\documentclass[amsmath,amssymb,showkeys,superscriptaddress,aps,prd]{revtex4}
\usepackage{graphicx,epstopdf}
\usepackage{subfig}
\def\xslash{x\!\!\!\slash }

\def\vel{\left|}
\def\ver{\right|}
\usepackage{colortbl}
\usepackage{xcolor}
\usepackage[colorlinks=true, urlcolor=blue, linkcolor=blue, citecolor=blue]{hyperref}

\begin{document}

\title{Magnetic and quadrupole moments of the $Z_c(3900)$}

\author{U.~\"{O}zdem}%
\email[]{uozdem@dogus.edu.tr}
\author{K.~Azizi}%
\email[]{kazizi@dogus.edu.tr}
\affiliation{Department of Physics, Do\v{g}u\c{s} University, Ac{\i}badem-Kad{\i}k\"{o}y, 34722 
\.{I}stanbul, T\"{u}rkiye}

\date{\today}
 
\begin{abstract}
The electromagnetic properties of the tetraquark state $Z_c(3900)$  are investigated in the diquark-antidiquark picture 
and its magnetic and quadrupole moments are extracted. To this end, the light-cone QCD sum rule 
in electromagnetic background field is used.
The magnetic and quadrupole  moments encode the spatial distributions of the charge and magnetization in the particle.
The result obtained for the magnetic moment is quite large and can be measured in future experiments.
We obtain a nonzero but small value for the quadrupole moment of $Z_c(3900)$
indicating a nonspherical charge distribution.
\end{abstract}

\maketitle

\section{Introduction}

In the conventional quark model, the predicted particles are mesons($q\bar q$), 
baryons($qqq$) and antibaryons($\bar q\bar q\bar q$). 
Hundreds of meson and baryon resonances have been observed till now. 
However, the quark model as well as  QCD as theory of 
strong interaction does not exclude
the existence of nonconventional particles. Hence, 
physicists have thought that there may be particles in  
different structures~\cite{GellMann:1964nj,Jaffe:1976ih,Witten:1979kh}.
Particles having different quark and gluon contents such as tetraquarks, pentaquarks,  
hybrids, glueballs and so on are called exotic states.
To explore the underlying structures of these states, 
many exotic structures have been suggested~[for instance, see~\cite{Nielsen:2009uh, Swanson:2006st, 
Voloshin:2007dx,Klempt:2007cp, Godfrey:2008nc,Faccini:2012pj,Esposito:2014rxa}].
Although predicted in the 1970s, there was not significant experimental evidence of their existence until recently.
Experimentally, the adventure of exotic states began when X(3872) 
was discovered by the Belle Collaboration~\cite{Choi:2003ue} 
and continued with the discovery of the Y(4260)
by the BABAr Collaboration~\cite{Aubert:2005rm}.
At present, more than twenty exotic states have been 
discovered in many experiments, 
most of which have been classified as  the XYZ family~(for details, see~\cite{Liu:2013waa}). 
The XYZ family has some decay channels that severely violate the isospin symmetry 
and negatively affect the identification 
of conventional charmonium/bottomonium states.
Because of that these newly observed XYZ states provide a good platform for studying the nonperturbative behavior of QCD.
The study of the properties of these particles is one of the most active and interesting branches of particle physics.

One of the most prominent particles among the exotic states is the charged $Z_{c}(3900)$ tetraquark.
The $Z_{c}^{\pm }(3900)$ state discovered by BESIII in the process 
$e^{+}e^{-}\rightarrow \pi ^{\pm}J/\psi$ \cite{Ablikim:2013mio}
 with a mass $3899.0 \pm 3.6 \pm 4.9$ MeV and width $\Gamma = 46 \pm 10 \pm 20 MeV$. 
 Almost at the same time this state was confirmed by the Belle Collaboration~\cite{Liu:2013dau},
 with a mass $3894.5 \pm 6.6 \pm 4.5$ MeV and width $\Gamma =63 \pm 24 \pm 26$ MeV. 
Its existence was also confirmed in Ref. \cite{Xiao:2013iha} on the basis of
the CLEO-c data analysis, with mass $3886.0 \pm 4.0 \pm 2.0$ MeV and width $\Gamma= 37 \pm 4 \pm 8 $ MeV.
The decays into $\pi ^{\pm}J/\psi$, reveal that $Z_{c}^{\pm }(3900)$
must be a tetraquark state with constituents $c\bar{c}u\bar{d}$ or $c\bar{c}d\bar{u}$~\cite{Braaten:2014qka}.
 Since the mass of $Z_{c}^{\pm }(3900)$ is very close to $X(3872)$, 
 it can be advised as the charged partner of the $X(3872)$ in a 
 tetraquark scenario.
 The properties of the $Z_{c}^{\pm }(3900)$ particle have 
 been investigated with different 
  theoretical models and approaches~\cite{Dias:2013xfa,Wang:2010rt,Wang:2013vex,Deng:2014gqa,Wang:2013daa,
Wilbring:2013cha,Dong:2013iqa,Ke:2013gia,Guo:2013sya,Gutsche:2014zda,Esposito:2014hsa,Goerke:2016hxf,
Agaev:2016dev, Agaev:2017tzv}.
Although the spectroscopic properties of these particles have been studied adequately, 
the internal structure and nature of the $ X(3872) $ and 
$ Z_c ^ \pm (3900) $ 
particles have not been fully understood yet. 
 For this reason, it is important to study their decay properties as well as  their interactions with other particles.
 In this context, examining the interaction of 
these particles with the photon can play an important role in understanding of their nature and internal structure.

A detailed study of the electromagnetic structures, such as electromagnetic multipole moments 
and electromagnetic form factors, of hadrons not only provides 
important information about the nonperturbative nature of QCD but also the multipole moments of the hadrons
are  important tools for understanding their internal structures in terms of quarks and gluons as well as their
geometric shape.
The electromagnetic multipole moments encode the spatial distributions of charge and magnetization
in the particle. 
In hadrons, quarks are the carriers of the charge, 
and thus these observables are directly connected to the spatial
distribution of quarks in hadrons, as well as a probe of the underlying dynamics.
The examination of the spatial distri
butions of the charge and magnetism carried by nuclei started in the 1950s.
The electromagnetic properties of the nucleon have been studied in the past extensively from unpolarized electron
scattering experiments-for reviews on experimental
progress, see for instance Refs.~\cite{Arrington:2011kb,Perdrisat:2006hj,Arrington:2006zm,HydeWright:2004gh,Gao:2003ag}.

There are many studies in the literature devoted to investigation of the multipole moments of the standard hadrons. 
However, unfortunately, almost nothing is known about the multipole moments of exotic particles 
and more detailed analyses are needed in this regard.
Since direct experimental information on the electromagnetic multipole moments of the exotic particles is very limited, 
theoretical studies can play an important role in this respect.
In this study, the tetraquark state $Z_c(3900)$  is investigated in the diquark-antidiquark picture 
and its magnetic and quadrupole moments are extracted. 
This is the first theoretical attempt to calculate the electromagnetic multipole moments
of the hidden-charm tetraquark states. 
To study the electromagnetic multipole moments, a nonperturbative method is needed. 
The light-cone QCD sum rule (LCSR) is one of the nonperturbative 
methods that has been successfully applied to study many 
nonperturbative properties of hadrons for decades~\cite{Chernyak:1990ag, Braun:1988qv, Balitsky:1989ry}. 
In the LCSR, the features of the particles under study are described in terms of the vacuum condensates and 
the light-cone distribution amplitudes (DAs).
Hence, any uncertainty in these parameters affects the estimations on the magnetic and quadrupole moments.

The rest of the paper is organized as follows: In Sec. II, the LCSR for the magnetic and quadrupole
moments of the $Z_c(3900)$ are derived. Section III is devoted to
the numerical analysis of the obtained sum rules. 
Section IV includes our concluding remarks. 
The explicit expressions of the photon distribution amplitudes, magnetic and quadrupole moments as well as
some details about calculations are moved to Appendixes A-C.

\section{Formalism}

In order to calculate the magnetic and quadrupole moments of the $Z_{c}(3900)$ state
in the framework of LCSR, we start from the correlation function
\begin{equation}
 \label{edmn01}
\Pi _{\mu \nu }(q)=i\int d^{4}xe^{ip\cdot x}\langle 0|\mathcal{T}\{J_{\mu}^{Z_c}(x)
J_{\nu }^{Z_c\dagger }(0)\}|0\rangle_{\gamma}, 
\end{equation}%
where $\gamma$ is the external electromagnetic field 
and $J_{\mu}$ is the interpolating current of the $Z_c(3900)$ state with quantum numbers $J^{PC}=1^{+-}$
in the diquark-antidiquark picture. It is given as
\begin{eqnarray}
J_{\mu }^{Z_{c}}(x) &=&\frac{i\epsilon \tilde{\epsilon}}{\sqrt{2}}\left\{ %
\left[ u_{a}^{T}(x)C\gamma _{5}c_{b}(x)\right] \left[ \overline{d}%
_{d}(x)\gamma _{\mu }C\overline{c}_{e}^{T}(x)\right] -\left[ u_{a}^{T}(x)C\gamma _{\mu }c_{b}(x)\right] \left[ \overline{%
d}_{d}(x)\gamma _{5}C\overline{c}_{e}^{T}(x)\right] \right\},
\label{eq:Curr}
\end{eqnarray}%
where $\epsilon =\epsilon _{abc}$, $\tilde{\epsilon}=\epsilon _{dec}$, $C$ is 
the charge conjugation matrix and $a,b,c,d,e$
are color indices.

We start to calculate the correlation function in terms of the hadronic parameters called the hadronic side.
To this end, we insert complete sets of intermediate states
having the same quantum numbers as the interpolating current of $Z_c(3900)$ into the correlation function, 
and isolate the contribution of the ground state. As a result the following expression is obtained: 
\begin{align}
\label{edmn04}
\Pi_{\mu\nu}^{Had} (p,q) = {\frac{\langle 0 \mid J_\mu^{Z_c} \mid
Z_c(p) \rangle}{p^2 - m_{Z_c}^2}} \langle Z_c(p) \mid Z_c(p+q) \rangle_\gamma
\frac{\langle Z_c(p+q) \mid {J^\dagger}_\nu^{Z_c} \mid 0 \rangle}{(p+q)^2 - m_{Z_c}^2} + \cdots,
\end{align}
where dots represent the contributions coming from the higher states and
continuum and $q$ is the momentum of the photon. The matrix element
$\langle 0 \mid J_\mu^{Z_c} \mid Z_c \rangle$ is parametrized as
\begin{align}
\label{edmn05}
\langle 0 \mid J_\mu^{Z_c} \mid Z_c \rangle = \lambda_{Z_c} \varepsilon_\mu^\theta\,,
\end{align}
with $\lambda_{Z_c}$ being the current coupling constant or residue of the $Z_c(3900)$ state.

In the presence of the electromagnetic background field, 
the vertex of the two axial vector mesons can be written in terms of form factors as follows~\cite{Brodsky:1992px}:
\begin{align}
\label{edmn06}
\langle Z_c(p,\varepsilon^\theta) \mid  Z_c (p+q,\varepsilon^{\delta})\rangle_\gamma
 &= - \varepsilon^\tau (\varepsilon^{\theta})^\alpha
(\varepsilon^{\delta})^\beta
\Bigg[ G_1(Q^2)~ (2p+q)_\tau ~g_{\alpha\beta}  +
G_2(Q^2)~ ( g_{\tau\beta}~ q_\alpha -  g_{\tau\alpha}~ q_\beta) \nonumber\\
&- \frac{1}{2 m_{Z_c}^2} G_3(Q^2)~ (2p+q)_\tau ~q_\alpha q_\beta  \Bigg]\,,
\end{align}
where $\varepsilon^\delta$ and $\varepsilon^{\theta}$ are the 
polarization vectors of the initial and final $Z_c(3900)$
mesons and $\varepsilon^\tau$ is the polarization vector of the photon. 
The form factors $G_1(Q^2)$, $G_2(Q^2)$  and $G_3(Q^2)$ can be written in terms of the charge
$F_C(Q^2)$,
magnetic $F_M(Q^2)$ and quadrupole $F_{\cal D}(Q^2)$ form factors in the following way: 
\begin{align}
\label{edmn07}
&F_C(Q^2) = G_1(Q^2) + \frac{2}{3} \lambda F_{\cal D}(Q^2)\,,\nonumber \\
&F_M(Q^2) = G_2(Q^2)\,,\nonumber \\
&F_{\cal D}(Q^2) = G_1(Q^2)-G_2(Q^2)+(1+\lambda) G_3(Q^2)\,,
\end{align}
where $\lambda=Q^2/4 m_{Z_c}^2$ with $Q^2=-q^2$. At $Q^2 = 0 $, the form
factors $F_C(Q^2=0)$, $F_M(Q^2=0)$, and $F_{\cal D}(Q^2=0)$ are related to the
electric charge, magnetic moment $\mu$, and quadrupole moment ${\cal D}$ in
the following way:
\begin{align}
\label{edmn08}
&e F_C(0) = e \,, \nonumber\\
&e F_M(0) = 2 m_{Z_c} \mu \,, \nonumber\\
&e F_{\cal D}(0) = m_{Z_c}^2 {\cal D}\,.
\end{align}
Using Eqs. (\ref{edmn04})-(\ref{edmn06}) and 
imposing the condition, $q\!\cdot\!\varepsilon = 0$, and performing
summation over polarization vectors, the correlation function takes the
form,
\begin{align}
\label{edmn09}
 \Pi_{\mu\nu}^{Had} &= \lambda_{Z_c}^2  \frac{\varepsilon^\tau}{ [m_{Z_c}^2 - (p+q)^2][m_{Z_c}^2 - p^2]}
 \Bigg[2 p_\tau F_C(0) \Bigg(g_{\mu\nu} -\frac{p_\mu
q_\nu-p_\nu q_\mu}{ m_{Z_c}^2 } \Bigg) \nonumber \\
&+ F_M (0) \Bigg(q_\mu g_{\nu\tau} - q_\nu g_{\mu\tau} +
\frac{1}{m_{Z_c}^2} p_\tau (p_\mu q_\nu - p_\nu q_\mu ) \Bigg)
- \Bigg(F_C(0) + F_{\cal D}(0)\Bigg) {\frac{p_\tau}{m_{Z_c}^2} } q_\mu
q_\nu \Bigg]\,.
\end{align}

The next step is to calculate the correlation function in  Eq.~(\ref{edmn01}) in terms of quarks and gluon
properties in the deep Euclidean region called the QCD side. For this aim, the interpolating currents are inserted 
into the correlation function and after the contracting of quark pairs using the Wick
theorem the following result is obtained:
\begin{eqnarray}
\label{edmn11}
&&\Pi _{\mu \nu }^{\mathrm{QCD}}(q)=-i\frac{\epsilon
\tilde{\epsilon}\epsilon^{\prime }\tilde{\epsilon}^{\prime }}{2}
\int d^{4}xe^{ipx} \langle 0 | \Bigg\{  
\mathrm{Tr}\Big[\gamma _{5}\widetilde{S}_{u}^{aa^{\prime }}(x)\gamma _{5}S_{c}^{bb^{\prime }}(x)\Big]
\mathrm{Tr}\Big[\gamma _{\mu }\widetilde{S}_{c}^{e^{\prime }e}(-x)\gamma _{\nu}S_{d}^{d^{\prime }d}(-x)\Big] \notag \\
&&-\mathrm{Tr}\Big[ \gamma _{\mu }\widetilde{S}_{c}^{e^{\prime}e}(-x)\gamma _{5}S_{d}^{d^{\prime }d}(-x)\Big] 
\mathrm{Tr}\Big[ \gamma_{\nu }\widetilde{S}_{u}^{aa^{\prime }}(x)\gamma _{5}S_{c}^{bb^{\prime }}(x)] \nonumber\\
&&-\mathrm{Tr}\Big[\gamma _{5}\widetilde{S}_{u}^{a^{\prime }a}(x)\gamma _{\mu }S_{c}^{b^{\prime}b}(x)\Big]  
\mathrm{Tr}\Big[ \gamma _{5}\widetilde{S}_{c}^{e^{\prime}e}(-x)\gamma _{\nu }S_{d}^{d^{\prime }d}(-x)\Big] \notag \\
&&+\mathrm{Tr}\Big[\gamma _{\nu }\widetilde{S}_{u}^{aa^{\prime }}(x)\gamma _{\mu }S_{c}^{bb^{\prime }}(x)\Big] 
\mathrm{Tr}\Big[\gamma _{5}\widetilde{S}_{c}^{e^{\prime }e}(-x)\gamma_{5}S_{d}^{d^{\prime }d}(-x)\Big]
 \Bigg\}| 0 \rangle_\gamma,
\end{eqnarray}%
where%
\begin{equation*}
\widetilde{S}_{c(q)}^{ij}(x)=CS_{c(q)}^{ij\mathrm{T}}(x)C,
\end{equation*}%
with $S_{q(c)}(x)$ being the quark propagators.
In the $x$-space for the light quark propagator we use in the $m_q\rightarrow 0$ limit
\begin{eqnarray}
\label{edmn12}
S_{q}(x)&=&i \frac{{\xslash}}{2\pi ^{2}x^{4}} 
- \frac{\bar qq}{12} - \frac{\bar qq}{192}
m_0^2 x^2  
-\frac {i g_s }{16 \pi^2 x^2} \int_0^1 dv~G^{\mu \nu} (vx) \Bigg[\rlap/{x} 
\sigma_{\mu \nu} 
+  \sigma_{\mu \nu} \rlap/{x}
 \Bigg].
\end{eqnarray}%

The heavy quark propagator is given, in terms of the second kind Bessel functions $K_{\nu }(x)$, as

\begin{eqnarray}
\label{edmn13}
&&S_{c}(x)=\frac{m_{c}^{2}}{4 \pi^{2}} \Bigg[ \frac{K_{1}(m_{c}\sqrt{-x^{2}}) }{\sqrt{-x^{2}}}
+i\frac{{\xslash}~K_{2}( m_{c}\sqrt{-x^{2}})}
{(\sqrt{-x^{2}})^{2}}\Bigg]
-\frac{g_{s}m_{c}}{16\pi ^{2}} \int_{0}^{1}dv~G^{\mu \nu }(vx)\Bigg[ (\sigma _{\mu \nu }{\xslash}
  +{\xslash}\sigma _{\mu \nu })\frac{K_{1}( m_{c}\sqrt{-x^{2}}) }{\sqrt{-x^{2}}}\nonumber\\
&&+2\sigma ^{\mu \nu }K_{0}( m_{c}\sqrt{-x^{2}})\Bigg].
\end{eqnarray}%

The correlation function contains  different types of contributions.
In the first part, one of the free quark propagators in Eq.~(\ref{edmn11}) is replaced by
\begin{align}
S^{free} \rightarrow \int d^4y\, S^{free} (x-y)\,\rlap/{\!A}(y)\, S^{free} (y)\,,
\end{align}
with $S^{free}$ representing the first term of the light or heavy quark propagators
and the remaining three propagators with the full quark propagators.
In the calculations the Fock-Schwinger gauge, $x_\mu A^\mu =0$, is used.

In the second case one of the light quark propagators in Eq.~(\ref{edmn11}) is replaced by
\begin{align}
\label{edmn14}
S_{\alpha\beta}^{ab} \rightarrow -\frac{1}{4} (\bar{q}^a \Gamma_i q^b)(\Gamma_i)_{\alpha\beta},
\end{align}
and the remaining propagators with the full quark propagators.
 Here, $\Gamma_i$ are the full set of Dirac matrices. Once 
Eq. (\ref{edmn14}) is plugged into Eq. (\ref{edmn11}), there appear matrix
elements such as $\langle \gamma(q)\vel \bar{q}(x) \Gamma_i q(0) \ver 0\rangle$
and $\langle \gamma(q)\vel \bar{q}(x) \Gamma_i G_{\alpha\beta}q(0) \ver 0\rangle$,
representing the nonperturbative contributions. 
These matrix elements can be expressed in terms of photon wave functions with definite
twists. Additionally, in principle, nonlocal operators such as
$\bar{q} G^2 q$ and $\bar{q}q\bar{q}q$ are anticipated to appear.
In this study, we take into account operators with
only one gluon field and contributions coming from three particle nonlocal operators 
and neglect terms with two gluons $\bar{q} G^2 q$, and four quarks $\bar{q}q\bar{q}q$. 
The matrix elements $\langle \gamma(q)\vel \bar{q}(x) \Gamma_i q(0) \ver 0\rangle$  
and $\langle \gamma(q)\vel \bar{q}(x) \Gamma_i G_{\alpha\beta}q(0) \ver 0\rangle$
are expressed in terms of the photon distribution amplitudes whose expressions
are given in Appendix A.
The QCD side of the correlation function can be obtained in terms of quarks and gluon properties 
 using Eqs.~(\ref{edmn11})-(\ref{edmn14}) and after performing the Fourier transformation to 
transfer the calculations to the momentum space.

The sum rules are obtained by matching the expression of the correlation function in terms
of quark-gluon properties to its expression in terms of the hadron properties, using their spectral
representation. 
In order to eliminate the subtraction terms in the spectral representation of the correlation function, the Borel
transformation with respect to the variables $p^2$ and $(p + q)^2$ is carried out. 
After the transformation, contributions from the excited and continuum states are also
exponentially suppressed. 
Finally, we choose the structures  $q_\mu \varepsilon_\nu$ and  
$(\varepsilon.p) q_\mu q_\nu$, respectively 
for the magnetic and quadrupole moments and obtain 
\begin{align}
 &\mu =\frac{e^{m_{Z_c}^2/M^2}}{\lambda_{Z_c}^2}\Bigg[\Pi_1+\Pi_2\Bigg],\nonumber\\
 &\mathcal{ D} = m^2_{Z_c}\frac{e^{m_{Z_c}^2/M^2}}{\lambda_{Z_c}^2}\Bigg[\Pi_3+\Pi_4\Bigg],
\end{align}
where the functions $ \Pi_1 $ and $ \Pi_3 $  indicate that one of the 
quark propagators enters the perturbative interaction with the photon and 
the remaining three propagators are taken as full propagators.
The functions $ \Pi_2 $ and $ \Pi_4 $ show that one of the light quark propagators 
enters the nonperturbative interaction with the photon and 
the remaining three propagators are taken as full propagators.
Explicit expressions of the $\Pi_1$, $\Pi_2$, $\Pi_3$  
and $\Pi_4$ are given in Appendix B.
As an example we show some details of the calculations i.e., Fourier and Borel transformations as well as 
the continuum subtraction, for a specific term in Appendix C.

\section{Numerical analysis}

In this section, we numerically analyze the results of calculations for magnetic and quadrupole moments.
We use $m_{Z_c}= 3899\pm 8.5~MeV$, 
$f_{3\gamma}=-0.0039~GeV^2$~\cite{Ball:2002ps}, $\overline{m}_c(m_c) = (1.275\pm 0.025)\,GeV$, 
$\langle \bar uu\rangle(1\,GeV) = 
\langle \bar dd\rangle(1\,GeV) =(-0.24\pm0.01)^3\,GeV^3$ \cite{Ioffe:2005ym},
$m_0^{2} = 0.8 \pm 0.1~GeV^2$, $\langle g_s^2G^2\rangle = 0.88~ GeV^4$~\cite{Nielsen:2009uh} and 
$\lambda_{Z_c}=m_{Z_c}f_{Z_c}=(1.79 \pm 0.12)\times 10^{-2}~GeV^5$~\cite{Agaev:2016dev,Agaev:2017tzv}.
We also need the value of the magnetic susceptibility which is obtained in different studies as 
$\chi(1\,GeV)=-2.85 \pm 0.5~GeV^{-2}$~\cite{Rohrwild:2007yt}, 
$\chi(1\,GeV)=-3.15 \pm 0.3~GeV^{-2}$~\cite{Ball:2002ps} and $\chi(1\,GeV)=-4.4~GeV^{-2}$~\cite{Belyaev:1984ic}. 
The parameters used in the photon distribution amplitudes are also given in Appendix A.

The predictions for the magnetic and quadrupole moments depend on two
 auxiliary parameters; the Borel mass parameter $M^2$ and continuum threshold $s_0$. According to the standard prescriptions in the method used the predictions should  weakly depend on these helping parameters.
 The continuum threshold represents the scale at which, the excited states and continuum start to
contribute to the correlation function.
Our analyses show that the results depend very weakly on $s_0$ in the interval
$(m_{Z_c}+0.3)^2~GeV^2 \leq s_0 \leq (m_{Z_c}+0.7)^2~GeV^2$.
%
%
The working region for $M^2$ is determined requiring that the contributions 
of the higher states and continuum are effectively suppressed.
In technique language, the upper bound on $M^2$ is found demanding the maximum pole contribution. 
The lower bound is obtained demanding that the contribution of the 
perturbative part exceeds 
the nonperturbative one and  series of the operator product expansion in the obtained sum rules converge.
The above requirements restrict the working
region of the Borel parameter to $5~GeV^2 \leq M^2 \leq 7~GeV^2$. It is worth nothing that with these intervals of $s_0$ and $M^2$ we receive a $ (85-93)\%  $ pole contribution, which nicely satisfies the requirements of the QCD sum rule approach.

 In Fig. 1, we plot the dependencies of the magnetic and 
 quadrupole moments on $M^2$ 
 at several fixed values of the continuum threshold $s_0$.
 As is seen, the variation of the results with respect to 
 the Borel parameters is considerable, 
but there is much less dependence of the quantities under 
consideration on the continuum threshold in its working interval. 
\begin{figure}
\centering
 \includegraphics[width=0.4\textwidth]{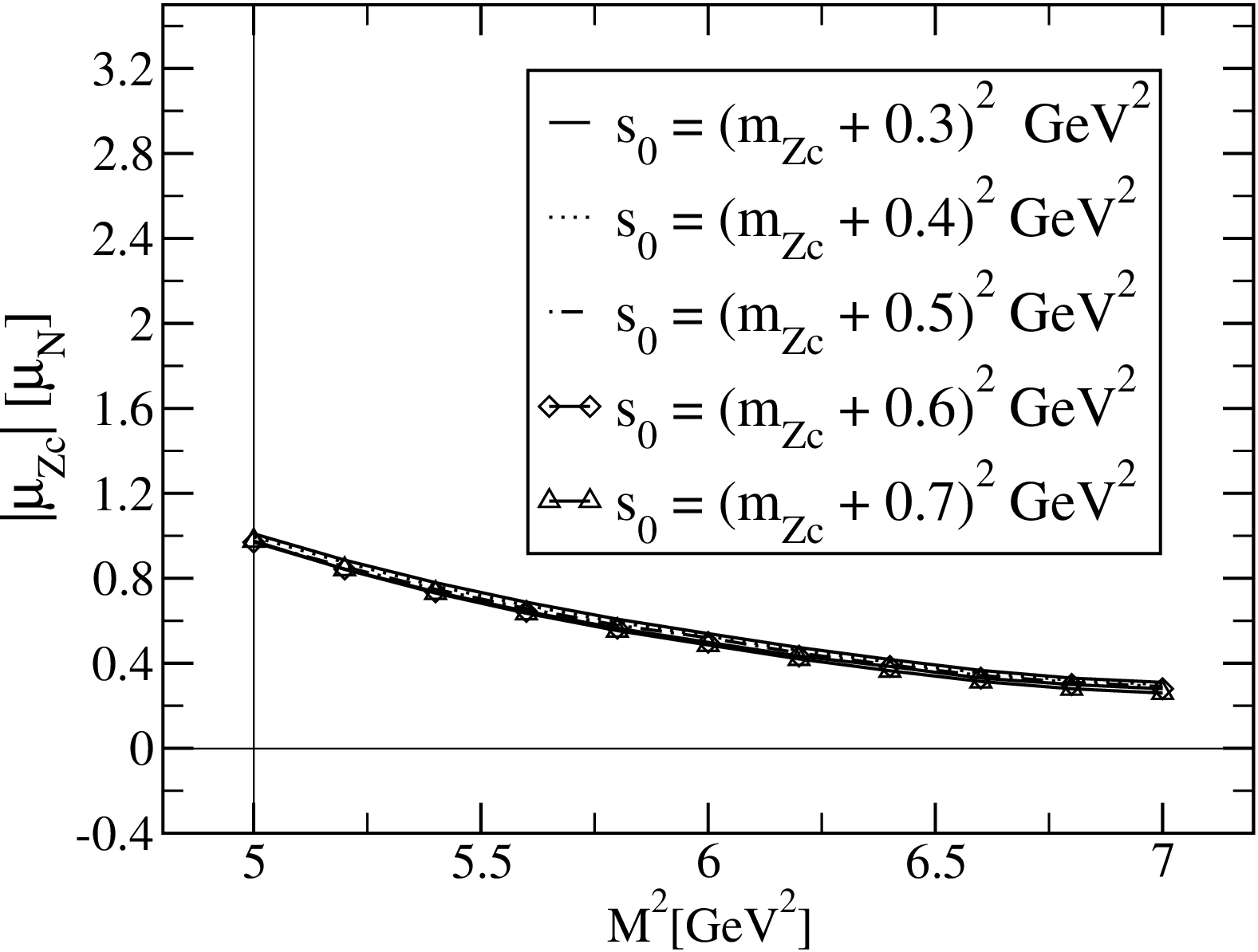}
 \hspace*{0.9cm}
 \includegraphics[width=0.4\textwidth]{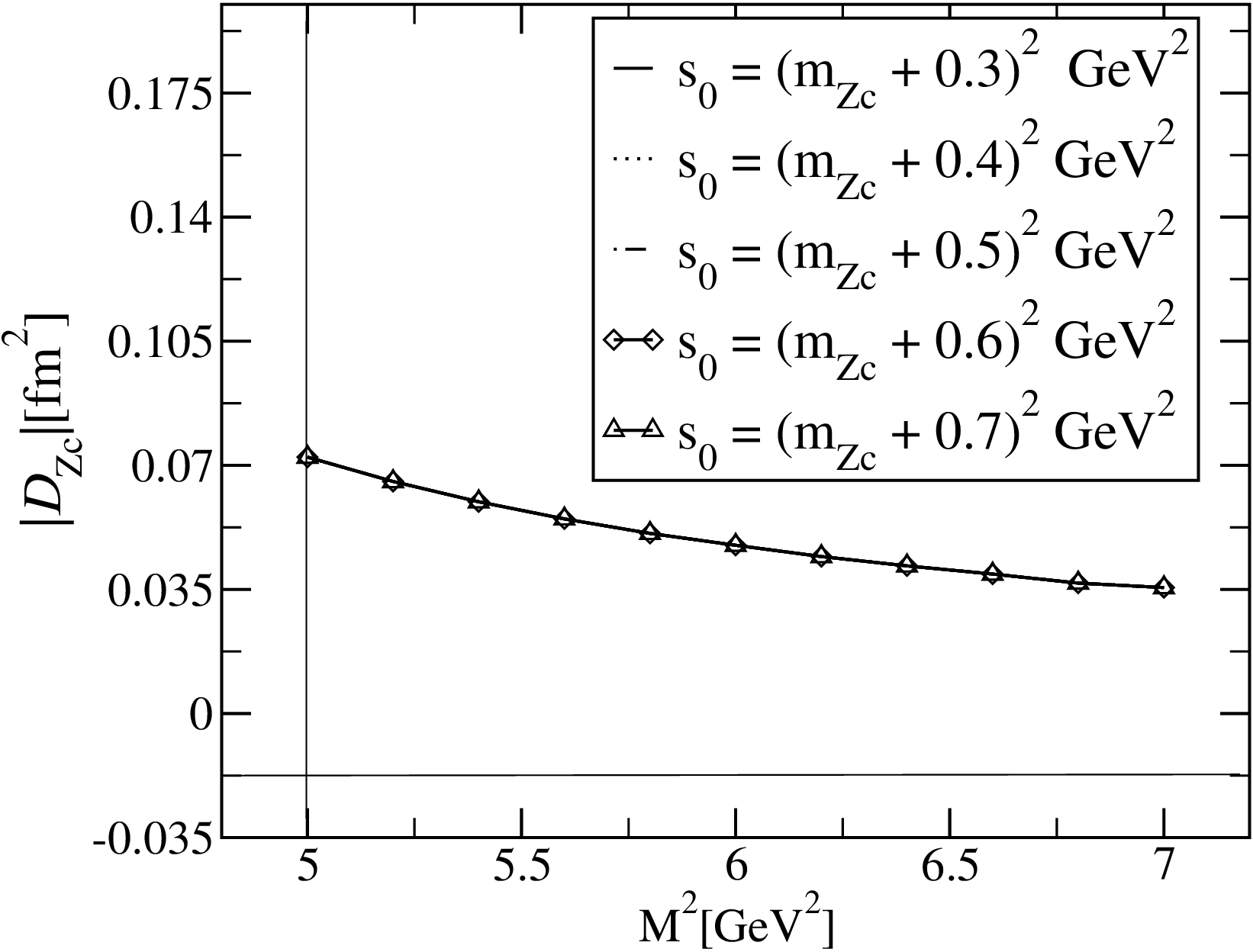}
 \caption{ The dependence of the magnetic and quadrupole moments; on the Borel parameter squared $M^{2}$
 at different fixed values of the continuum threshold.}
  \end{figure}
 In Fig. 2, we show the contributions of $ \Pi_1 $, $ \Pi_2 $, $ \Pi_3 $ and $ \Pi_4 $ functions
 to the results obtained at the average value of $s_0$ with respect to the Borel mass parameter.
 In the case of the magnetic moment, we see that the contribution of 
 $ \Pi_1 $ is the dominant contribution.
 $ \Pi_1 $ corresponds to roughly 65\% of the result in average, while the remaining 35\% belongs to $ \Pi_2 $.
 %
 In the case of quadrupole moment, we see that all contributions come 
 from $ \Pi_4 $ and the contribution of $ \Pi_3 $ is 0.

Our final results for the magnetic and quadrupole moments are 

\begin{align}
 &|\mu_{Z_c}| =0.67 \pm 0.32 ~\mu_N\nonumber\\
 &|\mathcal{D}_{Z_c}| =0.054 \pm 0.018 ~fm^2,
\end{align}
where the errors in the results come from the variations in the  calculations of the working regions of
 $M^2$ and $s_0$ as well as the uncertainties 
in the values of the input parameters and the photon DAs. 
We remark that the main source of uncertainties is the variations with respect to $M^{2}$ and the results very weakly depend on the choices of the continuum threshold.

\section{Discussion and concluding remarks}

We calculated the magnetic and quadrupole moments of the $Z_c(3900)$ state 
within the framework of the LCSR method. 
We obtained a measurable value for the magnetic dipole moment but a small value for the quadrupole 
moment indicating a nonspherical charge distribution.
It is useful to note that the values of the magnetic and quadrupole moments do not depend on the values of the
magnetic susceptibility $ \chi $  presented in the previous section.
It is worth mentioning also that there are different Lorentz structures to calculate 
the magnetic moment in the correlation function, but our result is almost independent 
of these structures. 
Any experimental measurements of the electromagnetic 
multipole moments of the $Z_c(3900)$
state and comparison of the obtained results with the
predictions of the present study may serve as
valuable knowledge on the internal structure of 
the tetraquark states as well as the nonperturbative nature of the QCD.
A comparison of our results on the electromagnetic 
multipole moments of the $Z_c(3900)$
state with those that can be obtained via considering 
different internal structures and interpolating currents, 
such as a molecular type one, would be very helpful in the 
determination of the internal structure of this multiquark state. A comparison of the results obtained  with the predictions of other approaches, such as lattice QCD, chiral perturbation theory,
quark model, etc., would also be interesting.

 \begin{figure}
\centering
 \includegraphics[width=0.4\textwidth]{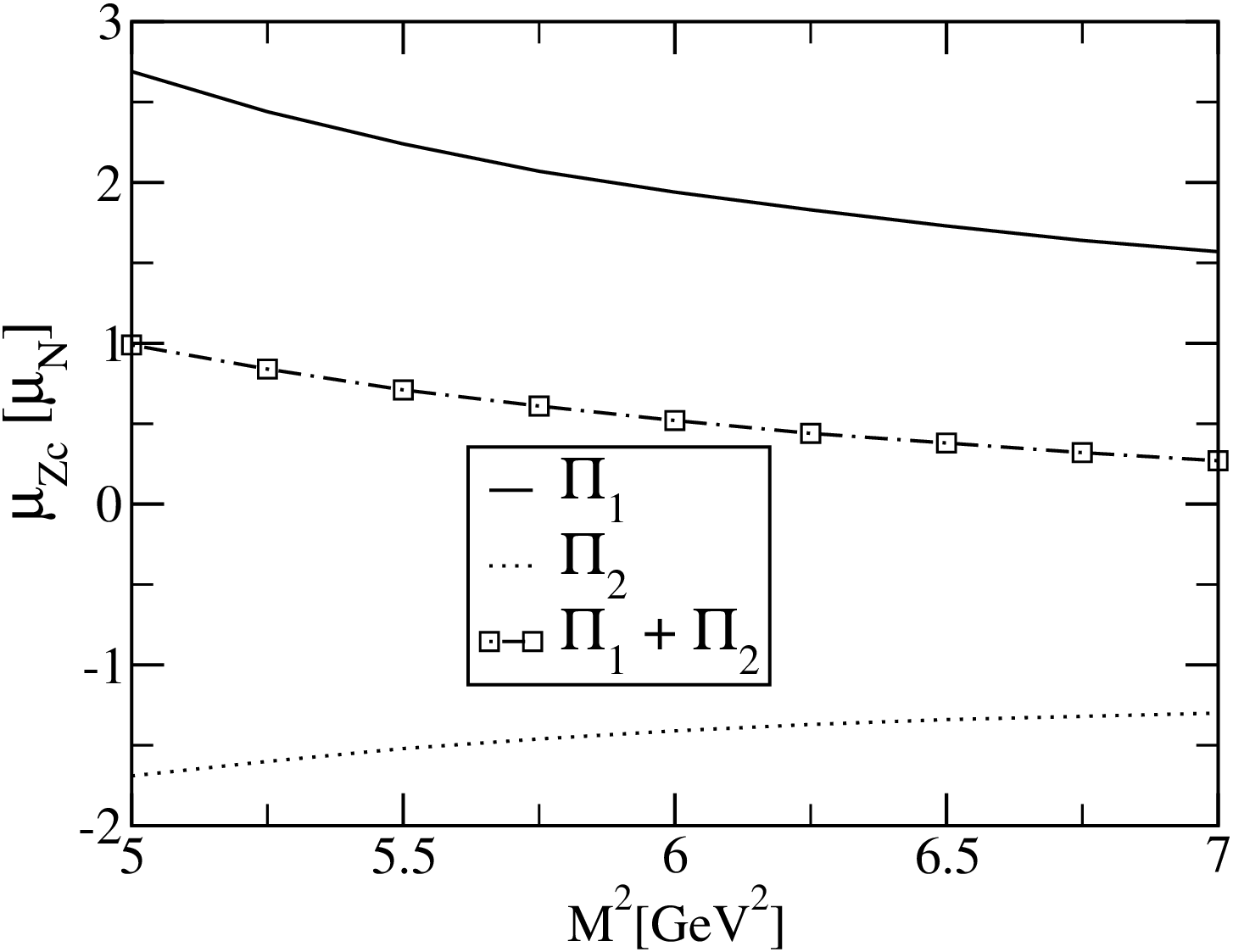}
 \hspace*{0.7cm}
 \includegraphics[width=0.4\textwidth]{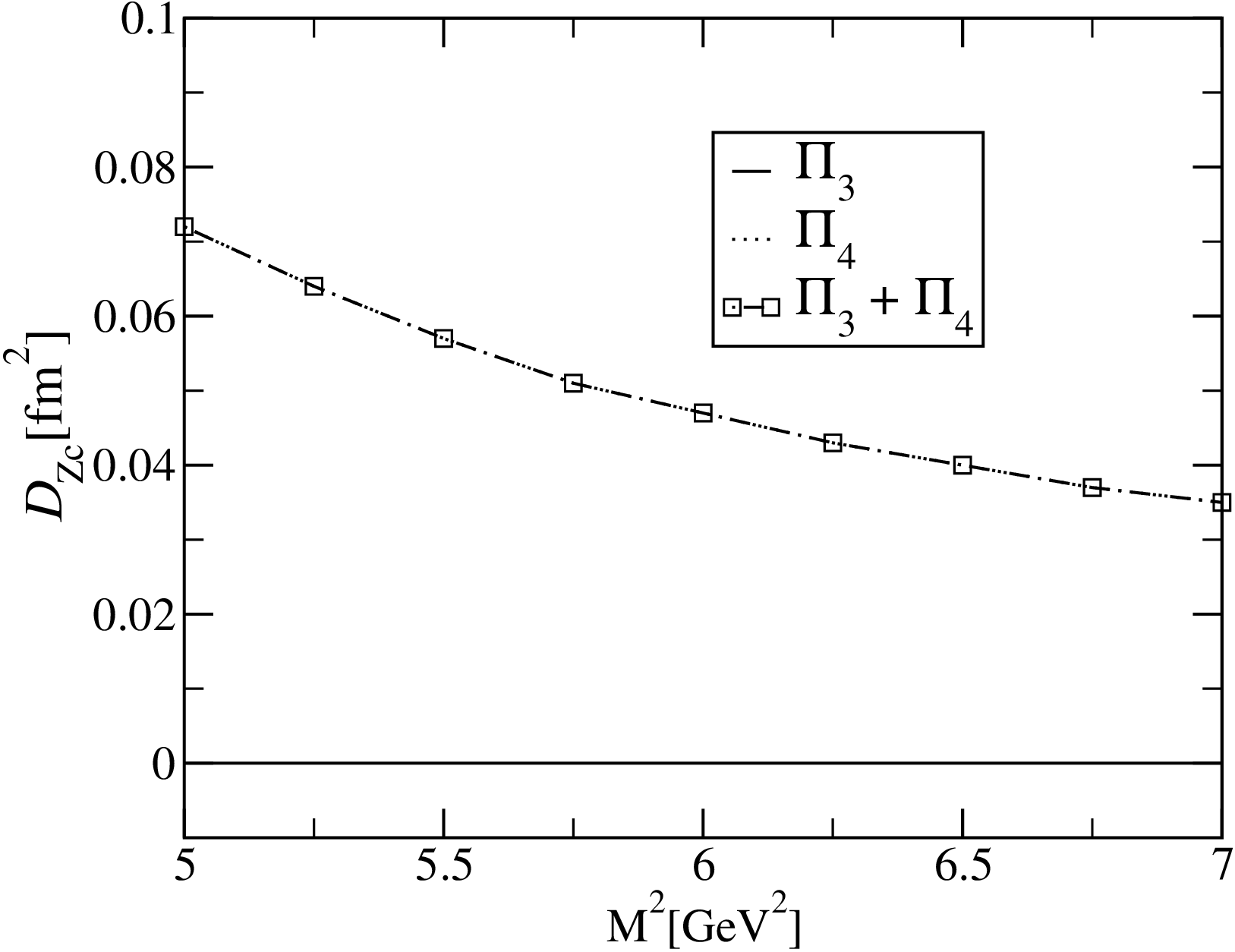}
 \caption{ Comparison of the contributions to the magnetic and quadrupole moments with respect to $M^2$
at average value of $s_0$.}
\end{figure}

\section{Acknowledgements}

This work has been supported by the Scientific and
Technological Research Council of Turkey (T\"{U}B\.{I}TAK)
under the Grant No. 115F183.

\newpage
\appendix
\section*{Appendix A: Photon distribution amplitudes}
		        In this appendix, we present the definitions of the matrix elements of the
form $\langle \gamma(q)\vel \bar{q}(x) \Gamma_i q(0) \ver 0\rangle$  
and $\langle \gamma(q)\vel \bar{q}(x) \Gamma_i G_{\mu\nu}q(0) \ver 0\rangle$ in terms of the photon
DAs, and the explicit expressions of the 
photon distribution amplitudes \cite{Ball:2002ps},

\begin{eqnarray*}
\label{esbs14}
&&\langle \gamma(q) \vert  \bar q(x) \gamma_\mu q(0) \vert 0 \rangle
= e_q f_{3 \gamma} \left(\varepsilon_\mu - q_\mu \frac{\varepsilon
x}{q x} \right) \int_0^1 du e^{i \bar u q x} \psi^v(u)
\nonumber \\
&&\langle \gamma(q) \vert \bar q(x) \gamma_\mu \gamma_5 q(0) \vert 0
\rangle  = - \frac{1}{4} e_q f_{3 \gamma} \epsilon_{\mu \nu \alpha
\beta } \varepsilon^\nu q^\alpha x^\beta \int_0^1 du e^{i \bar u q
x} \psi^a(u)
\nonumber \\
&&\langle \gamma(q) \vert  \bar q(x) \sigma_{\mu \nu} q(0) \vert  0
\rangle  = -i e_q \langle \bar q q \rangle (\varepsilon_\mu q_\nu - \varepsilon_\nu
q_\mu) \int_0^1 du e^{i \bar u qx} \left(\chi \varphi_\gamma(u) +
\frac{x^2}{16} \mathbb{A}  (u) \right) \nonumber \\ 
&&-\frac{i}{2(qx)}  e_q \bar qq \left[x_\nu \left(\varepsilon_\mu - q_\mu
\frac{\varepsilon x}{qx}\right) - x_\mu \left(\varepsilon_\nu -
q_\nu \frac{\varepsilon x}{q x}\right) \right] \int_0^1 du e^{i \bar
u q x} h_\gamma(u)
\nonumber \\
&&\langle \gamma(q) | \bar q(x) g_s G_{\mu \nu} (v x) q(0) \vert 0
\rangle = -i e_q \langle \bar q q \rangle \left(\varepsilon_\mu q_\nu - \varepsilon_\nu
q_\mu \right) \int {\cal D}\alpha_i e^{i (\alpha_{\bar q} + v
\alpha_g) q x} {\cal S}(\alpha_i)
\nonumber \\
&&\langle \gamma(q) | \bar q(x) g_s \tilde G_{\mu \nu}(v
x) i \gamma_5  q(0) \vert 0 \rangle = -i e_q \langle \bar q q \rangle \left(\varepsilon_\mu q_\nu -
\varepsilon_\nu q_\mu \right) \int {\cal D}\alpha_i e^{i
(\alpha_{\bar q} + v \alpha_g) q x} \tilde {\cal S}(\alpha_i)
\nonumber \\
&&\langle \gamma(q) \vert \bar q(x) g_s \tilde G_{\mu \nu}(v x)
\gamma_\alpha \gamma_5 q(0) \vert 0 \rangle = e_q f_{3 \gamma}
q_\alpha (\varepsilon_\mu q_\nu - \varepsilon_\nu q_\mu) \int {\cal
D}\alpha_i e^{i (\alpha_{\bar q} + v \alpha_g) q x} {\cal
A}(\alpha_i)
\nonumber \\
&&\langle \gamma(q) \vert \bar q(x) g_s G_{\mu \nu}(v x) i
\gamma_\alpha q(0) \vert 0 \rangle = e_q f_{3 \gamma} q_\alpha
(\varepsilon_\mu q_\nu - \varepsilon_\nu q_\mu) \int {\cal
D}\alpha_i e^{i (\alpha_{\bar q} + v \alpha_g) q x} {\cal
V}(\alpha_i) \nonumber\\
&& \langle \gamma(q) \vert \bar q(x)
\sigma_{\alpha \beta} g_s G_{\mu \nu}(v x) q(0) \vert 0 \rangle  =
e_q \langle \bar q q \rangle \left\{
        \left[\left(\varepsilon_\mu - q_\mu \frac{\varepsilon x}{q x}\right)\left(g_{\alpha \nu} -
        \frac{1}{qx} (q_\alpha x_\nu + q_\nu x_\alpha)\right) \right. \right. q_\beta
\nonumber \\ && -
         \left(\varepsilon_\mu - q_\mu \frac{\varepsilon x}{q x}\right)\left(g_{\beta \nu} -
        \frac{1}{qx} (q_\beta x_\nu + q_\nu x_\beta)\right) q_\alpha
-
         \left(\varepsilon_\nu - q_\nu \frac{\varepsilon x}{q x}\right)\left(g_{\alpha \mu} -
        \frac{1}{qx} (q_\alpha x_\mu + q_\mu x_\alpha)\right) q_\beta
\nonumber \\ &&+
         \left. \left(\varepsilon_\nu - q_\nu \frac{\varepsilon x}{q.x}\right)\left( g_{\beta \mu} -
        \frac{1}{qx} (q_\beta x_\mu + q_\mu x_\beta)\right) q_\alpha \right]
   \int {\cal D}\alpha_i e^{i (\alpha_{\bar q} + v \alpha_g) qx} {\cal T}_1(\alpha_i)
\nonumber \\ &&+
        \left[\left(\varepsilon_\alpha - q_\alpha \frac{\varepsilon x}{qx}\right)
        \left(g_{\mu \beta} - \frac{1}{qx}(q_\mu x_\beta + q_\beta x_\mu)\right) \right. q_\nu
\nonumber \\ &&-
         \left(\varepsilon_\alpha - q_\alpha \frac{\varepsilon x}{qx}\right)
        \left(g_{\nu \beta} - \frac{1}{qx}(q_\nu x_\beta + q_\beta x_\nu)\right)  q_\mu
\nonumber \\ && -
         \left(\varepsilon_\beta - q_\beta \frac{\varepsilon x}{qx}\right)
        \left(g_{\mu \alpha} - \frac{1}{qx}(q_\mu x_\alpha + q_\alpha x_\mu)\right) q_\nu
\nonumber \\ &&+
         \left. \left(\varepsilon_\beta - q_\beta \frac{\varepsilon x}{qx}\right)
        \left(g_{\nu \alpha} - \frac{1}{qx}(q_\nu x_\alpha + q_\alpha x_\nu) \right) q_\mu
        \right]      
    \int {\cal D} \alpha_i e^{i (\alpha_{\bar q} + v \alpha_g) qx} {\cal T}_2(\alpha_i)
\nonumber \\
&&+\frac{1}{qx} (q_\mu x_\nu - q_\nu x_\mu)
        (\varepsilon_\alpha q_\beta - \varepsilon_\beta q_\alpha)
    \int {\cal D} \alpha_i e^{i (\alpha_{\bar q} + v \alpha_g) qx} {\cal T}_3(\alpha_i)
\nonumber \\ &&+
        \left. \frac{1}{qx} (q_\alpha x_\beta - q_\beta x_\alpha)
        (\varepsilon_\mu q_\nu - \varepsilon_\nu q_\mu)
    \int {\cal D} \alpha_i e^{i (\alpha_{\bar q} + v \alpha_g) qx} {\cal T}_4(\alpha_i)
                        \right\}~,
\end{eqnarray*}
where $\varphi_\gamma(u)$ is the leading twist-2, $\psi^v(u)$,
$\psi^a(u)$, ${\cal A}(\alpha_i)$ and ${\cal V}(\alpha_i)$, are the twist-3, and
$h_\gamma(u)$, $\mathbb{A}(u)$, ${\cal S}(\alpha_i)$, ${\cal{\tilde S}}(\alpha_i)$, ${\cal T}_1(\alpha_i)$, ${\cal T}_2(\alpha_i)$, ${\cal T}_3(\alpha_i)$ 
and ${\cal T}_4(\alpha_i)$ are the
twist-4 photon DAs.
The measure ${\cal D} \alpha_i$ is defined as
\begin{eqnarray*}
\label{nolabel05}
\int {\cal D} \alpha_i = \int_0^1 d \alpha_{\bar q} \int_0^1 d
\alpha_q \int_0^1 d \alpha_g \delta(1-\alpha_{\bar
q}-\alpha_q-\alpha_g)~.\nonumber
\end{eqnarray*}

The expressions of the DAs entering into the above matrix elements are
defined as:

\begin{eqnarray}
\varphi_\gamma(u) &=& 6 u \bar u \left( 1 + \varphi_2(\mu)
C_2^{\frac{3}{2}}(u - \bar u) \right),
\nonumber \\
\psi^v(u) &=& 3 \left(3 (2 u - 1)^2 -1 \right)+\frac{3}{64} \left(15
w^V_\gamma - 5 w^A_\gamma\right)
                        \left(3 - 30 (2 u - 1)^2 + 35 (2 u -1)^4
                        \right),
\nonumber \\
\psi^a(u) &=& \left(1- (2 u -1)^2\right)\left(5 (2 u -1)^2 -1\right)
\frac{5}{2}
    \left(1 + \frac{9}{16} w^V_\gamma - \frac{3}{16} w^A_\gamma
    \right),
\nonumber \\
h_\gamma(u) &=& - 10 \left(1 + 2 \kappa^+\right) C_2^{\frac{1}{2}}(u
- \bar u),
\nonumber \\
\mathbb{A}(u) &=& 40 u^2 \bar u^2 \left(3 \kappa - \kappa^+
+1\right)  +
        8 (\zeta_2^+ - 3 \zeta_2) \left[u \bar u (2 + 13 u \bar u) \right.
\nonumber \\ && + \left.
                2 u^3 (10 -15 u + 6 u^2) \ln(u) + 2 \bar u^3 (10 - 15 \bar u + 6 \bar u^2)
        \ln(\bar u) \right],
\nonumber \\
{\cal A}(\alpha_i) &=& 360 \alpha_q \alpha_{\bar q} \alpha_g^2
        \left(1 + w^A_\gamma \frac{1}{2} (7 \alpha_g - 3)\right),
\nonumber \\
{\cal V}(\alpha_i) &=& 540 w^V_\gamma (\alpha_q - \alpha_{\bar q})
\alpha_q \alpha_{\bar q}
                \alpha_g^2,
\nonumber \\
{\cal T}_1(\alpha_i) &=& -120 (3 \zeta_2 + \zeta_2^+)(\alpha_{\bar
q} - \alpha_q)
        \alpha_{\bar q} \alpha_q \alpha_g,
\nonumber \\
{\cal T}_2(\alpha_i) &=& 30 \alpha_g^2 (\alpha_{\bar q} - \alpha_q)
    \left((\kappa - \kappa^+) + (\zeta_1 - \zeta_1^+)(1 - 2\alpha_g) +
    \zeta_2 (3 - 4 \alpha_g)\right),
\nonumber \\
{\cal T}_3(\alpha_i) &=& - 120 (3 \zeta_2 - \zeta_2^+)(\alpha_{\bar
q} -\alpha_q)
        \alpha_{\bar q} \alpha_q \alpha_g,
\nonumber \\
{\cal T}_4(\alpha_i) &=& 30 \alpha_g^2 (\alpha_{\bar q} - \alpha_q)
    \left((\kappa + \kappa^+) + (\zeta_1 + \zeta_1^+)(1 - 2\alpha_g) +
    \zeta_2 (3 - 4 \alpha_g)\right),\nonumber \\
{\cal S}(\alpha_i) &=& 30\alpha_g^2\{(\kappa +
\kappa^+)(1-\alpha_g)+(\zeta_1 + \zeta_1^+)(1 - \alpha_g)(1 -
2\alpha_g)\nonumber +\zeta_2[3 (\alpha_{\bar q} - \alpha_q)^2-\alpha_g(1 - \alpha_g)]\},\nonumber \\
\tilde {\cal S}(\alpha_i) &=&-30\alpha_g^2\{(\kappa -\kappa^+)(1-\alpha_g)+(\zeta_1 - \zeta_1^+)(1 - \alpha_g)(1 -
2\alpha_g)\nonumber +\zeta_2 [3 (\alpha_{\bar q} -\alpha_q)^2-\alpha_g(1 - \alpha_g)]\}.
\end{eqnarray}

Numerical values of parameters used in DAs; $\varphi_2(1~GeV) = 0$, 
$w^V_\gamma = 3.8 \pm 1.8$, $w^A_\gamma = -2.1 \pm 1.0$, 
$\kappa = 0.2$, $\kappa^+ = 0$, $\zeta_1 = 0.4$, $\zeta_2 = 0.3$, 
$\zeta_1^+ = 0$, and $\zeta_2^+ = 0$.
\newpage
\section*{Appendix B:}
In this appendix, we present the explicit expressions for the functions, $\Pi_1$, $\Pi_2$, $\Pi_3$ and $\Pi_4$:

\begin{align}
 \Pi_1 &= \frac{3 m_c^4 M^2}{64 \pi^6}\Bigg[2(3e_u+4e_d-2e_c)N[3,3,0]-3m_c(e_u+e_d-e_c)N[3,4,1]
 -e_c\Big(8N[4,2,0]-m_cN[5,2,1]\Big)\Bigg]\nonumber\\
&-\frac{ m_c^3 M^2 \langle g_s^2G^2\rangle \langle \bar qq \rangle}{12288 \pi^4}\Big(3e_u+3e_d-2e_c\Big)
N[1,2,1]\nonumber\\
&+\frac{ m_c^4 M^2 \langle g_s^2G^2\rangle }{147456 \pi^6}\Big(2e_u+2e_d-e_c\Big)
\Big(N[1,3,1]+N[2,2,1]\Big)\nonumber\\
&-\frac{ m_c^3 M^2 \langle g_s^2G^2\rangle }{18432 \pi^6}\Big(2e_u+2e_d-e_c\Big)N[1,2,0]\nonumber\\
&-\frac{m_c^2 M^2 \langle g_s^2 G^2 \rangle }{1536 \pi^6}\Big(e_u+e_d-7e_c\Big)N[2,2,0]\nonumber\\
&+\frac{m_c^3 M^2}{24576 \pi^6}\Bigg[-(17 e_u+17e_d-31e_c) \langle g_s^2 G^2 \rangle 
+576(3e_u+3e_d-4e_c)\pi^2 m_c \langle \bar qq \rangle\Bigg]N[2,3,1] \nonumber\\
&+\frac{m_c^2}{13824 M^6 \pi^6}\Bigg[(e_u+e_d+e_c)\langle g_s^2G^2\rangle M^2
-(e_u+e_d)~36 \pi^2 m_c \langle \bar qq\rangle (3m_0^2+16 M^2) \Bigg]\Bigg(64~m_c^6FlP[-3,4,0]
\nonumber\\
&-48 m_c^4 FlP[-2,4,0]+12~m_c^2FlP[-1,4,0]-FlP[0,4,0]\Bigg)
\nonumber\\
&-\frac{m_c \langle \bar qq\rangle }{110592 M^8 \pi^4}\Bigg[
3(e_u+e_d)\langle g_s^2G^2\rangle M^2 (3m_0^2+16 M^2)
+2e_c\Big(\langle g_s^2G^2\rangle M^2 (3m_0^2-4 M^2)
+18 \pi^2 m_c \langle \bar qq\rangle \nonumber\\
&(3m_0^4-128 M^4)\Big)\Bigg]\Big(16~m_c^4FlP[-1,2,0]-8m_c^2~FlP[0,2,0]+FlP[1,2,0]\Big)\nonumber\\
&+ \frac{ e_c m_c m_0^2 \langle \bar qq\rangle^2 }{221184 M^6 \pi^4}
\Big(5 \langle g_s^2G^2\rangle+1152 \pi^2 m_c \langle \bar qq\rangle \Big)\Bigg[16~m_c^4 FlP[1,2,1]
-8m_c^2 FlP[2,2,1]-FlP[3,2,1]\Bigg]\nonumber\\
&+\frac{ e_c m_c^3 m_0^2 \langle \bar qq\rangle }{48 M^4 \pi^4}\Bigg[
64m_c^4FlP[-2,3,0]+28 m_c^2FlP[0,3,0]-5 FlP[1,3,0]\Bigg]\nonumber\\
&+\frac{ e_c m_c^2 m_0^4 \langle \bar qq\rangle^2 }{6144 M^8 \pi^4}\Bigg[
16~m_c^4 FlP[3,2,2]-8~m_c^2 FlP[4,2,2]+FlP[5,2,2]\Bigg].
\end{align}

\begin{align}
&\Pi_2 =\frac{ m_c^4 \langle \bar qq \rangle }{128 \pi^4}\Bigg[e_u\Big(WFD[\mathcal{S},\bar v]-2WF[\mathcal{S},\bar v]\Big)+
e_d\Big(WFD[\mathcal{S},v]-2WF[\mathcal{S},v]\Big) \Bigg]
\Big(4N[2,3,0]-M^2 N[2,3,1]\Big)\nonumber\\
 &+\frac{m_c^4 \langle \bar qq \rangle }{32 \pi^4}
 \Bigg[e_u\Big(-2 WF[\mathcal{T}_1,\bar v]-2 WF[\mathcal{T}_2,\bar v] + 2 WF[\tilde S,\bar v]+WFD[\mathcal{T}_1,\bar v]
 +WFD[\mathcal{T}_2,\bar v]- WFD[\tilde S,\bar v]\Big)\nonumber\\
 &+e_d\Big(-8 WF[\mathcal{T}_1,v]-2WF[\mathcal{T}_2,v]+2 WF[\mathcal{S},v]
 +4 WFD[\mathcal{T}_2,v]-WFD[\mathcal{S},v]-WFD[\tilde S,v]\Big) \Bigg]N[1,4,0]\nonumber\\
 &+\frac{m_c^4 M^2 \langle \bar qq \rangle }{256 \pi^4}
 \Bigg[ e_u\Big(4 WF[\mathcal{T}_1,\bar v] + 4 WF[\mathcal{T}_2,\bar v]-2 WF[\tilde S,\bar v]
 -2 WFD[\mathcal{T}_1,\bar v]-2 WFD[\mathcal{T}_2,\bar v]+ 2 WFD[\tilde S,v]\Big)\nonumber\\
 &+e_d\Big(13 WF[\mathcal{T}_1,v]+7 WF[\mathcal{T}_2,v]
 -WF[\tilde S,v]-8 WFD[\mathcal{T}_1,v]-2 WFD[\mathcal{T}_2,v] + 2 WFD[\tilde S,v]\Big)\Bigg]N[1,4,1]\nonumber\\
&+\frac{f_{3 \gamma}m_c^4}{64 \pi^4}\Bigg[16 (e_u-e_d) WFD[\psi^a,u]+e_uWFD[\mathcal{V},\bar v]
+e_dWFD[\mathcal{V},v]\Bigg]N[3,3,0]\nonumber\\
 &+\frac{ m_c^3 M^2}{512 \pi^4}\Bigg[
  m_c \langle \bar qq \rangle \Bigg\{ 
   e_u \Big(-2 WF[\mathcal{S},\bar v] + 6 WF[\mathcal{T}_1,\bar v] + 6 WF[\mathcal{T}_2,\bar v] - 2 WF[\tilde S,\bar v]
 +3 WFD[\mathcal{S},\bar v]\nonumber\\
 &- 3 WFD[\mathcal{T}_1,\bar v] - 3 WFD[\mathcal{T}_2,\bar v] + 3 WFD[\tilde S,\bar v] \Big)
 +e_d\Big(-2WF[\mathcal{S},v] + 18 WF[\mathcal{T}_1,v]
 + 12 WF[\mathcal{T}_2,v] + 4 WF[\tilde S,v] \nonumber\\
 &+3 WFD[\mathcal{S},v]- 12 WFD[\mathcal{T}_1,v]- 3 WFD[\mathcal{T}_2,v] + 3 WFD[\tilde S,v]\Big)\Bigg\}
 +2e_u f_{3\gamma} M^2\Big(2 WFD[\mathcal{A},\bar v] + WFD[\mathcal{V},\bar v]\Big)\nonumber\\
 &+2 e_d f_{3\gamma} M^2  \Big( 2 WFD[\mathcal{A},v]+3WFD[\mathcal{V},v]\Big)
  \Bigg]N[2,3,1]\nonumber\\
&+\frac{  m_c^2 M^2f_{3 \gamma}\langle g_s^2 G^2\rangle}{110592 \pi^4}
\Big[-10(e_u+e_d) \psi^a(u_0) +2(e_u-4e_d)\varphi_\gamma(u_0)
-5(e_u-e_d)WFD[\psi^a,u]+4(e_u-e_d)WFD[\psi^\nu,u]\Big]\nonumber\\
&N[1,1,0]\nonumber\\
 &+\frac{\langle \bar qq \rangle m_c^4 M^4}{2048 \pi^4}\Bigg[
 e_u\Big(- 2 WF[\mathcal{S},\bar v]+ 2 WF[\mathcal{T}_1,\bar v] + 2 WF[\mathcal{T}_2,v]- 2 WF[\tilde S,v]
 +WFD[\mathcal{S},\bar v] 
 - WFD[\mathcal{T}_1,\bar v]\nonumber\\
 &- WFD[\mathcal{T}_2,\bar v] + WFD[\tilde S,\bar v]\Big)
  +e_d\Big(-2 WF[\mathcal{S},v]+8 WF[\mathcal{T}_1,v]
 +  2 WF[\mathcal{T}_2,v]- 2 WF[\tilde S,v] + WFD[\mathcal{S},v]-\nonumber\\
 &4 WFD[\mathcal{T}_1,v]- WFD[\mathcal{T}_2,v]+ WFD[\tilde S,v]\Big)
 \Bigg]N[2,3,2]\nonumber\\
 &-\frac{f_{3 \gamma} m_c^4}{32 \pi^4}\Bigg[e_u WFD[\mathcal{V},\bar v]+e_d WFD[\mathcal{V},v]\Bigg]N[2,4,0]\nonumber\\
 &+\frac{ m_c^3}{128 \pi^4}\Bigg[m_c \langle \bar qq \rangle\Bigg\{
 e_u \Big(2 WF[\mathcal{S},\bar v] + 2 WF[\mathcal{T}_1,\bar v]
 - 2 WF[\mathcal{T}_2,\bar v] + 2 WF[\tilde S,\bar v] 
 - WFD[\mathcal{S},\bar v] +  WFD[\mathcal{T}_1,\bar v]  
 \nonumber\\
 & +  WFD[\mathcal{T}_2,\bar v]- WFD[\tilde S,\bar v]\Big)
  +e_d \Big(2 WF[\mathcal{S},v]-8 WF[\mathcal{T}_1,v]
 - 2 WF[\mathcal{T}_2, v] + 2 WF[\tilde S,v] - WFD[\mathcal{S},v] \nonumber\\
 &+ 4 WFD[\mathcal{T}_1, v]+ 2 WFD[\mathcal{T}_2,v] -  WFD[\tilde S, v]\Big)\Bigg\}  
 +f_{3 \gamma} M^2\Big(e_u WFD[\mathcal{V},\bar v] 
 + e_d WFD[\mathcal{V},v]\Big)\Bigg] N[2,3,0]\nonumber\\
 &- \frac{m_c^4 M^2 f_{3 \gamma} }{128 \pi^4}\Bigg(e_u WFD[\mathcal{V},\bar v] 
 + e_d WFD[\mathcal{V},v]\Bigg) N[2,4,1]\nonumber\\
 &+\frac{ m_c^2 M^4 \langle g_s^2 G^2\rangle }{1769472 \pi^4}
 \Bigg[-4(4e_u-e_d)\chi m_c \langle \bar qq \rangle WFD[\varphi_\gamma, u]
 -16(e_u-e_d) f_{3\gamma}WFD[\psi^a,u]
 -11e_u f_{3\gamma} WFD[\mathcal{A},\bar v]\nonumber\\
 & -11e_d f_{3\gamma} WFD[\mathcal{A}, v] \Bigg]
N[2,2,2]\nonumber
\end{align}

\begin{align}
 &+\frac{ m_c^3 }{6912 \pi^4}\Bigg[-54 M^2\langle \bar qq \rangle \Bigg \{
  e_u\Big( 2 WF[\mathcal{T}_1,\bar v] + 2 WF[\mathcal{T}_2,\bar v] 
 + 3 WF[\tilde S,\bar v] - WFD[\mathcal{T}_1,\bar v] - WFD[\mathcal{T}_2,\bar v] 
+ WFD[\tilde S,\bar v]\Big)\nonumber\\
 & +e_d\Big(5 WF[\mathcal{T}_1, v]  + 5 WF[\mathcal{T}_2, v]
 + WF[\tilde S, v] - 4 WFD[\mathcal{T}_1, v] - WFD[\mathcal{T}_2, v]+ WFD[\tilde S, v]\Big)\Bigg\} \nonumber\\
 &-54 m_c M^2 f_{3\gamma }\Big(e_u WFD[\mathcal{A}, \bar v]+ e_d WFD[\mathcal{A}, v]\Big)
 +(6e_u+7e_d)\chi \langle g_s^2 G^2 \rangle \langle \bar qq \rangle WF[\varphi_\gamma,u]
 \Bigg]N[1,3,0]\nonumber\\
&+ \frac{m_c^3 M^2 \chi \langle g_s^2 G^2 \rangle \langle \bar qq \rangle }{55296 \pi^4}
\Bigg[-13(e_u-e_d) \varphi_\gamma(u_0)-2(7e_u-6e_d)WFD[\varphi_\gamma,u]\Bigg]
N[1,3,1]\nonumber\\
&-\frac{ m_c^3 M^4 \chi \langle g_s^2 G^2 \rangle \langle \bar qq \rangle }{110592 \pi^4}
 \Bigg[(e_u-e_d) WFD[\varphi_\gamma,u]\Bigg]N[1,3,2]\nonumber\\
&+\frac{m_c^2 \langle g_s^2 G^2\rangle}{110592 \pi^4}\Bigg[52(e_u-e_d)M^2 \chi \langle \bar qq \rangle \varphi_\gamma(u_0)
-4 (8e_u+5e_d)M^2 \chi \langle \bar qq \rangle WFD[\varphi_\gamma,u] 
-30(e_u-e_d)m_c f_{3\gamma}WFD[\psi^a,u]\nonumber\\
&+11 \langle \bar qq \rangle \Bigg\{
e_u\Big(6WF[\mathcal{S},\bar v]-4WF[\mathcal{T}_1,\bar v]-10WF[\mathcal{T}_2,\bar v]+2WF[\mathcal{T}_3,\bar v]
-WF[\mathcal{T}_4,\bar v]+WF[\tilde S,\bar v]-3WFD[\mathcal{S},\bar v]\nonumber\\
&+2WFD[\mathcal{T}_1,\bar v]+5WFD[\mathcal{T}_2,\bar v]-WFD[\mathcal{T}_3,\bar v]
+WFD[\mathcal{T}_4,\bar v]-2WFD[\tilde S,\bar v]\Big)\nonumber\\
&+e_d\Big(6WF[\mathcal{S}, v]-4WF[\mathcal{T}_1, v]-10WF[\mathcal{T}_2, v]+2WF[\mathcal{T}_3, v]
-WF[\mathcal{T}_4,\bar v]+WF[\tilde S,\bar v]-3WFD[\mathcal{S},\bar v]\nonumber\\
&+2WFD[\mathcal{T}_1, v]+5WFD[\mathcal{T}_2, v]-WFD[\mathcal{T}_3, v]
+WFD[\mathcal{T}_4, v]-2WFD[\tilde S, v]\Big)
\Bigg]N[1,2,0]\nonumber\\
&+\frac{f_{3 \gamma} m_c^4 M^2}{512 \pi^4}\Bigg[3e_uWFD[\mathcal{V},\bar v]+3e_d WFD[\mathcal{V},v]
+16(e_u-e_d)WFD[\psi^a,u]\Bigg]N[3,3,1]\nonumber\\
&+\frac{m_c^2 M^2 \langle g_s^2 G^2\rangle }{884736 \pi^4}\Bigg[
40 m_c f_{3 \gamma} (e_u-e_d)\psi^a(u_0)-8f_{3 \gamma} m_c (e_u- e_d)\psi^\nu(u_0)
-16 m_c f_{3 \gamma} (4e_u-e_d)WF[\psi^\gamma,u]\nonumber\\
&+30f_{3 \gamma}m_c (e_u-e_d)WFD[\psi^a,u] -2(8e_u-5e_d)\langle \bar qq \rangle WFD[A,u]
+11 \langle \bar qq\rangle \Bigg\{ 
e_u\langle \bar qq\rangle \Big(-6WF[\mathcal{S},\bar v]+4WF[\mathcal{T}_1,\bar v]
\nonumber\\
&+10WF[\mathcal{T}_2,\bar v]-2WF[\mathcal{T}_3,\bar v]+4WF[\mathcal{T}_4,\bar v]-4WF[\tilde S,\bar v]
+3WFD[\mathcal{S},\bar v]-2WFD[\mathcal{T}_1,\bar v]-5WFD[\mathcal{T}_2,\bar v]\nonumber\\
&+WFD[\mathcal{T}_3,\bar v]
-2WFD[\mathcal{T}_4,\bar v]+2WFD[\tilde S,\bar v]\Big)
+e_d \Big(-6WF[\mathcal{S},v]+4WF[\mathcal{T}_1,v]
+10WF[\mathcal{T}_2,v]-2WF[\mathcal{T}_3,v]\nonumber\\
&+4WF[\mathcal{T}_4,v]-4WF[\tilde S,v]
+3WFD[\mathcal{S},v]-2WFD[\mathcal{T}_1,v]-5WFD[\mathcal{T}_2,v]+WFD[\mathcal{T}_3,v]-2WFD[\mathcal{T}_4,v]\nonumber\\
&+2WFD[\tilde S,v]\Big)\Bigg\}
\Bigg]N[1,2,1]\nonumber\\
&+\frac{m_c^2 \langle g_s^2 G^2\rangle}{110592 \pi^4}\Bigg[ 
-104(e_u-e_d)m_c \chi \langle g_s^2 G^2\rangle \varphi_\gamma(u_0)
+e_u f_{3\gamma}(11\langle g_s^2 G^2\rangle+432 m_c^2 M^2 )WFD[\mathcal{A},\bar v]\nonumber\\
&+e_d f_{3\gamma}(11\langle g_s^2 G^2\rangle+432 m_c^2 M^2 )WFD[\mathcal{A}, v]
+4(28e_u-19e_d)m_c \chi \langle g_s^2 G^2\rangle \langle \bar qq \rangle WFD[\varphi_\gamma,u]\nonumber\\
&+16(e_u-e_d)f_{3\gamma}\langle g_s^2 G^2\rangle WFD[\psi^a,u]
\Bigg]N[2,2,1]\nonumber\\
&+\frac{\langle g_s^2 G^2\rangle m_c^2 M^4}{28311552 \pi^4}\Bigg[11 e_d f_{3 \gamma} WFD[\mathcal{A},v]
-11 f_{3 \gamma} e_u WFD[\mathcal{A},\bar v]+ 4 \chi m_c \langle \bar qq\rangle(e_d-4e_u)
WFD[\varphi_\gamma,u]\nonumber\\
&-16 f_{3 \gamma} (e_u-e_d)WFD[\psi^a,u]\Bigg]N[2,2,2]\nonumber\\
&-\frac{m_c^2 M^4 f_{3\gamma} }{2048 \pi^4}\Bigg[e_uWFD[\mathcal{V},\bar v]+e_dWFD[\mathcal{V},v]
-16(e_u-e_d)WFD[\psi^a,u]\Bigg]N[3,3,2]\nonumber
 \end{align}

\begin{align}
&+\frac{m_c^2 M^4\langle g_s^2 G^2\rangle }{3538944 \pi^4}\Bigg[10(e_u-e_d)m_c f_{3\gamma}WFD[\psi^a,u]
+2(8e_u+5e_d)\langle \bar qq \rangle WFD[A,u]
+11e_u\Big(6WF[\mathcal{S},\bar v]-4WF[\mathcal{T}_1,\bar v]\nonumber\\
&-10WF[\mathcal{T}_2,\bar v]+2WF[\mathcal{T}_3,\bar v]-4WF[\mathcal{T}_4,\bar v]
+4WF[\tilde S,\bar v]-3WFD[\mathcal{S},\bar v]+2WFD[\mathcal{T}_1,\bar v]
+5WFD[\mathcal{T}_2,\bar v]\nonumber\\
&-3WFD[\mathcal{T}_3,\bar v]+2WFD[\mathcal{T}_4,\bar v]-WFD[\tilde S,\bar v]\Big)
+11e_d\Big(6WF[\mathcal{S}, v]-4WF[\mathcal{T}_1, v]-10WF[\mathcal{T}_2, v]+2WF[\mathcal{T}_3, v]\nonumber\\
&-4WF[\mathcal{T}_4, v]
+4WF[\tilde S, v]-3WFD[\mathcal{S}, v]+2WFD[\mathcal{T}_1, v]
+5WFD[\mathcal{T}_2, v]-3WFD[\mathcal{T}_3, v]+2WFD[\mathcal{T}_4, v]\nonumber\\
&-WFD[\tilde S, v]\Big)
\Bigg]N[1,2,2]\nonumber\\
&-\frac{m_c^2}{55296 \pi^4}\Bigg[f_{3\gamma}\Big(11\langle g_s^2 G^2\rangle+432m_c^2 M^2 \Big)
\Big(e_u WFD[\mathcal{A},\bar v]+e_d WFD[\mathcal{A}, v]\Big)
+12 m_c \langle \bar qq \rangle \Bigg\{ 36 M^2 \Big( e_u WFD[\mathcal{S},\bar v]\nonumber\\
&+e_dWFD[\mathcal{S},v]\Big)-(4e_u-3e_d)\chi \langle g_s^2 G^2\rangle WFD[\varphi_\gamma,u]\Bigg\}
+16(e_u-e_d)f_{3\gamma}\langle g_s^2 G^2\rangle WFD[\psi^a,u]\Bigg]N[2,2,0]
\nonumber\\
&+\frac{m_c \langle \bar qq \rangle^2 }{1990656 M^{10} \pi^2}\Bigg[
-(e_u-e_d)\Big(5 \langle g_s^2 G^2\rangle(23m_0^2-8M^2)-1728 m_c^2 m_0^2 M^2\Big)A(u_0)\nonumber\\
&+4m_0^2 \langle g_s^2 G^2\rangle \Big(10(e_u-e_d)M^2 \chi \varphi_\gamma(u_0)
+(7e_u+2e_d)WF[h_\gamma,u]\Big)\Bigg]\Bigg(16m_c^4FLNP[2,3,2]-8m_c^2FlNP[3,3,2]\nonumber\\
&-FlNP[4,3,2]\nonumber\\
&+\frac{m_c m_0^2 \langle g_s^2 G^2\rangle \langle \bar qq \rangle}{165888 M^8 \pi^2}
(e_u-e_d)\Bigg(4m_c^2FlNP[4,1,2]-FlNP[5,1,2]\Bigg)A(u_0)\nonumber\\
&-\frac{f_{3\gamma} m_0^2 \langle g_s^2 G^2\rangle \langle \bar qq \rangle}{73728 M^{8} \pi^2}
(e_u-e_d)\Bigg(16m_c^4FlNP[3,3,2]-8m_c^2FlNP[4,2,2]+FlNP[5,2,2]\Bigg)\psi^a(u_0)\nonumber\\
&+\frac{5 m_c m_0^2 \langle g_s^2 G^2\rangle \langle \bar qq \rangle^2}{995328 M^{10} \pi^2}
(e_u-e_d)\Bigg(16m_c^4FlNP[4,3,3]-8m_c^2FlNP[5,3,3]+FlNP[6,3,3]\Bigg)A(u_0)\nonumber\\
&+\frac{m_c m_0^2 \langle \bar qq \rangle}{9216 M^8 \pi^2}\Bigg[-4(e_u-e_d)m_c \psi^a(u_0)
-\langle \bar qq \rangle\Bigg\{ e_u\Big(4WF[\mathcal{T}_1,\bar v]+WF[\mathcal{T}_2,\bar v]\Big)
+e_d\Big(4WF[\mathcal{T}_1, v]+WF[\mathcal{T}_2, v]\Big)\Big\}\Bigg]\nonumber\\
&\Bigg(64m_c^6FlNP[1,4,2]-48m_c^4FlNP[2,4,2]+12m_c^2FlNP[3,4,2]-FlNP[4,4,2]\Bigg)\nonumber\\
&+\frac{m_c \langle g_s^2G^2 \rangle \langle \bar qq \rangle^2}{82944 M^8 \pi^2}\Bigg[
(e_u+e_d)(3m_0^2-2M^2)A(u_0)
-m_0^2\Big(2M^2 \chi \varphi_\gamma(u_0)+WF[h_\gamma,u]\Big)\Bigg]\Bigg(4m_c^2FlNP[2,1,1]\nonumber\\
&-FlNP[3,1,1]\Bigg)\nonumber\\
&+\frac{m_c^5 \langle \bar qq \rangle }{124416 M^{10}\pi^4}\Bigg[\Bigg\{e_u\Bigg(
-3456\pi^2 m_c^2 M^2 \langle \bar qq \rangle (3m_0^2-2M^2)
+\langle g_s^2G^2 \rangle\Big(-69m_c M^4+20 \pi^2\langle \bar qq \rangle(16m_0^2-11M^2)\Big)\Bigg)\nonumber\\
&+e_d\Bigg(
3456\pi^2 m_c^2 M^2 \langle \bar qq \rangle (3m_0^2-2M^2)
+5\langle g_s^2G^2 \rangle\Big(9m_c M^4+4 \pi^2\langle \bar qq \rangle(-16m_0^2+11M^2)\Big)\Bigg)\Bigg\}A(u_0)\nonumber\\
&+4\pi^2\langle \bar qq \rangle\Bigg\{-(e_u-e_d)M^2 \chi \Big(5\langle g_s^2G^2 \rangle(11m_0^2-8M^2)
-1728m_c^2m_0^2M^2\Big)\varphi_\gamma(u_0)
+4e_u\Big(216 m_c^2 m_0^2 M^2\nonumber\\
&+7\langle g_s^2G^2 \rangle(-2m_0^2+M^2)\Big)
+4e_d\Big(-216 m_c^2 m_0^2 M^2+\langle g_s^2G^2 \rangle(-4m_0^2+2M^2)\Big)\Bigg\}WF[h_\gamma,u]\Bigg]
FlNP[0,3,1]\nonumber
\end{align}

\begin{align}
& +\frac{\langle \bar qq \rangle }{663552 M^8 \pi^2}\Bigg[
 9(e_u-e_d)f_{3\gamma}\langle g_s^2 G^2\rangle (5m_0^2-4M^2)\psi^a(u_0)
 +2(4e_u-e_d)m_0^2 f_{3\gamma}\langle g_s^2 G^2 \rangle WF[\psi^\nu,u]\nonumber\\
 &+(e_u-e_d)m_0^2 f_{3\gamma}\langle g_s^2 G^2 \rangle \psi^\nu(u_0)
 +m_c m_0^2 M^2 \langle \bar qq \rangle \Bigg\{ e_u \Big(3WF[\mathcal{S},\bar v]-2WF[\mathcal{T}_1, \bar v]
 -2WF[\mathcal{T}_2, \bar v]+2WF[\mathcal{T}_3, \bar v]\nonumber\\
 &+2WF[\mathcal{T}_4, \bar v]-2WF[\tilde S, \bar v]\Big)
 + e_d \Big(3WF[\mathcal{S}, v]-2WF[\mathcal{T}_1,  v]
 -2WF[\mathcal{T}_2,  v]+2WF[\mathcal{T}_3,  v]+2WF[\mathcal{T}_4, v]\nonumber\\
 &-2WF[\tilde S, v]\Big)\Bigg\}\Bigg] \Bigg(16m_c^4FlNP[1,2,1]-8m_c^2FlNP[2,2,1]
 +FlNP[3,2,1]\Bigg)\nonumber\\
&+ \frac{1}{1327104 M^8 \pi^4}\Bigg[27(e_u-e_d)M^4 \langle g_s^2 G^2\rangle \langle g\bar qq\rangle A(u_0)
 +3(e_u-e_d)f_{3\gamma}\langle g_s^2 G^2\rangle\Big( 5m_cM^4+6\pi^2\langle \bar qq \rangle
 (-m_0^2+4M^2)\Big)\psi^a(u_0)\nonumber\\
 &-\pi^2 f_{3\gamma}\langle g_s^2 G^2\rangle\langle \bar qq\rangle (3m_0^2-4M^2)\Big(
 2(e_u-4e_d)\psi^\nu(u_0)+4(4e_u-e_d)WF[\psi^\nu,u]\Big)\nonumber\\
 &+\langle \bar qq\rangle\Bigg\{
 \Big(-23 M^2 \langle g_s^2 G^2\rangle-864 \pi^2 m_c \langle \bar qq\rangle(m_0^2-2M^2)\Big)
 \Big(  e_u WF[\mathcal{S},\bar v]+e_dWF[\mathcal{S}, v]\Big)\nonumber\\
 &+\Big(17 M^2 \langle g_s^2 G^2\rangle+288 \pi^2 m_c \langle \bar qq\rangle(m_0^2-2M^2)\Big)
 \Big(  e_u WF[\mathcal{T}_1,\bar v]+e_dWF[\mathcal{T}_1, v]\Big)\nonumber\\
&+ \Big(102 M^2 \langle g_s^2 G^2\rangle+1728 \pi^2 m_c \langle \bar qq\rangle(m_0^2-2M^2)\Big)
 \Big(  e_u WF[\mathcal{T}_2,\bar v]+e_dWF[\mathcal{T}_2, v]\Big)\nonumber\\
 &- \Big(36 M^2 \langle g_s^2 G^2\rangle+1728 \pi^2 m_c \langle \bar qq\rangle(m_0^2-2M^2)\Big)
 \Big(  e_u WF[\mathcal{T}_3,\bar v]+e_dWF[\mathcal{T}_3, v]\Big)\nonumber\\
 &- \Big(36 M^2 \langle g_s^2 G^2\rangle+1728 \pi^2 m_c \langle \bar qq\rangle(m_0^2-2M^2)\Big)
 \Big(  e_u WF[\tilde S,\bar v]+e_dWF[\tilde S, v]\Big)\Bigg\}\Bigg]\Bigg(16m_c^4FlNP[1,2,0]
  \nonumber\\
  &-8m_c^2FlNP[0,2,0]+FlNP[-1,2,0]\Bigg)\nonumber\\
   &+\frac{m_c}{15925248 M^{12} \pi^4}\Bigg[10368(e_d-e_u)m_c\langle \bar qq \rangle M^8 A(u_0) 
 +48 (e_d-e_u)f_{3\gamma}M^4\Big(7 \langle g_s^2G^2 \rangle M^2\nonumber\\
 &+144~m_c~\pi^2\langle \bar qq \rangle(4M^2-3m_0^2)\Big)\psi^a(u_0)
 -288(e_u-e_d)f_{3\gamma}M^4\Big(\langle g_s^2G^2 \rangle M^2
 +24~m_c~\pi^2\langle \bar qq \rangle(4M^2-3m_0^2)\Big)\psi^\nu(u_0)\nonumber\\
 &+2592 m_c~M^8\langle \bar qq\rangle \Big(e_d WF[\mathcal S,v]+e_u WF[\mathcal S,\bar v]\Big)
 -e_d \langle \bar qq\rangle M^4\Big(10368~m_c~M^4+33696 \pi^2 \langle \bar qq\rangle m_0^2\nonumber\\
 &+44928 \pi^2 \langle \bar qq\rangle M^2\Big)WF[\mathcal T_1,v]
 -e_u \langle \bar qq\rangle M^4\Big(2592~m_c~M^4+10368 \pi^2 \langle \bar qq\rangle m_0^2
 -13824 \pi^2 \langle \bar qq\rangle M^2\Big)WF[\mathcal T_1,\bar v]\nonumber\\
 &-e_d \langle \bar qq\rangle M^4\Big(2592~m_c~M^4+18144 \pi^2 \langle \bar qq\rangle m_0^2
 +24192 \pi^2 \langle \bar qq\rangle M^2\Big)WF[\mathcal T_2,v]
 -e_u \langle \bar qq\rangle M^4\Big(2592~m_c~M^4\nonumber\\
 &+10368 \pi^2 \langle \bar qq\rangle m_0^2
 -13824 \pi^2 \langle \bar qq\rangle M^2\Big)WF[\mathcal T_2,\bar v]
 -4320e_d \pi^2\langle \bar qq \rangle^2 M^4(3 m_0^2-4 M^2)WF[\tilde S,v]\nonumber\\
 &-1728 e_u \pi^2\langle \bar qq \rangle^2 M^4(3 m_0^2-4 M^2)WF[\tilde S,\bar v]
 +576 (e_d-e_u) f_{3\gamma}\langle g_s^2 G^2 \rangle M^6 WF[\psi^\nu,u]\nonumber\\
 &+13824 (e_u-e_d)\pi^2 f_{3\gamma}\langle \bar qq \rangle m_c ~ M^4 (3m_0^2-4 M^2)WF[\psi^\nu,u]\nonumber\\
& +(39e_u-38e_d)\langle g_s^2 G^2 \rangle \langle \bar qq \rangle m_c~M^4 WFD[\mathbb A,u]
+8(4e_u-e_d)\pi^2 \langle g_s^2 G^2 \rangle \langle \bar qq \rangle^2 (5m_0^2-4M^2)WFD[\mathbb A,u]\nonumber\\
&-e_d f_{3\gamma} M^4 \Big(204 \langle g_s^2 G^2 \rangle M^2
-2592 \pi^2 \langle \bar qq \rangle m_c~m_0^2+3456\pi^2\langle \bar qq \rangle m_c~M^2\Big)WFD[\mathcal A, v] 
-e_u f_{3\gamma} M^4 \Big(204 \langle g_s^2 G^2 \rangle M^2\nonumber\\
&-2592 \pi^2 \langle \bar qq \rangle m_c~m_0^2
+3456\pi^2\langle \bar qq \rangle m_c~M^2\Big)WFD[\mathcal A, \bar v] 
-e_d f_{3\gamma} M^4 \Big(138 \langle g_s^2 G^2 \rangle M^2
+2592 \pi^2 \langle \bar qq \rangle m_c~m_0^2\nonumber\\
&-3456\pi^2\langle \bar qq \rangle m_c~M^2\Big)WFD[\mathcal V, v]
-e_u f_{3\gamma} M^4 \Big(138 \langle g_s^2 G^2 \rangle M^2
+2592 \pi^2 \langle \bar qq \rangle m_c~m_0^2\nonumber\\
&-3456\pi^2\langle \bar qq \rangle m_c~M^2\Big)WFD[\mathcal V, \bar v]
+32(e_d-e_u) \pi^2 \chi \langle g_s^2 G^2 \rangle \langle \bar qq \rangle^2 M^2
(m_0^2-M^2)WFD[\varphi_\gamma,u]\Bigg]\nonumber\\
&\Big(FlNP[0,4,0]-8m_c^2 FlNP[1,4,0]+16 m_c^4 FlNP[2,4,0]\Big)\nonumber
\end{align}

\begin{align}
&+\frac{m_c \langle g_s^2 G^2 \rangle \langle \bar qq \rangle}{82944 M^8 \pi^2}\Bigg[
(e_u+e_d)\Big((3m_0^2-4M^2)A(u_0)+4M^2 \chi (-m_0^2+2M^2) \varphi_\gamma(u_0)
+(-3m_0^2+4M^2)WF[h_\gamma,u]\Big)\Bigg]\nonumber\\
&\Bigg(-4m_c^2FlNP[-1,1,0]+FlNP[0,1,0]\Bigg)\nonumber\\
  &+\frac{m_c m_0^2 \langle g_s^2 G^2\rangle \langle \bar qq \rangle^2 }{7962624 M^{12}\pi^4} \Bigg[
 -(4e_u-e_d)WFD[\mathcal{A},u]\Bigg]\Bigg(16m_c^4FlNP[2,4,2]-8m_c^2FlNP[3,4,2]+FlNP[4,4,2]\Bigg)\nonumber\\
 &+\frac{m_c^3 f_{3\gamma}}{92160 M^8 \pi^4}\Bigg[e_uWFD[\mathcal{V},\bar v]
 +e_dWF[\mathcal{V},v]\Bigg]\Bigg(64 m_c^6FlNP[4,6,0]-48m_c^4FlNP[3,6,0]+12m_c^2FlNP[2,6,0]\nonumber\\
 &-FlNP[1,6,0]\Bigg)\nonumber\\
 &-\frac{f_{3\gamma} m_0^2\langle \bar qq\rangle}{73728 M^8 \pi^2}\Bigg[
 e_d WFD[\mathcal{V},v]+e_u WFD[\mathcal{V},\bar v]\Bigg]FlNP[3,4,1]\nonumber\\
  &+\frac{m_c}{3981312 M^{12} \pi^4}\Bigg[-12(e_d-e_u)f_{3 \gamma} M^4 \Big( 7\langle g_s^2 G^2 \rangle
 +144\pi^2 \langle \bar qq\rangle m_c(4M^2-5m_0^2)\Big)\psi^a(u_0)
 \nonumber\\
 &+\langle \bar qq\rangle \pi^2\Bigg(1728(e_d-e_u)f_{3\gamma}m_0^2m_c~M^4 \psi^\nu(u_0)
 +216 e_d M^4\langle \bar qq\rangle (45 m_0^2-32 M^2)WF[\mathcal{T}_1,v]\nonumber\\
 &+864 e_u M^4\langle \bar qq\rangle (3m_0^2-2M^2) WF[\mathcal{T}_1,\bar v]
 +216 e_d M^4 \langle \bar qq\rangle (15 m_0^2-8M^2)WF[\mathcal{T}_2, v]\nonumber\\
 &+864 e_u M^4\langle \bar qq\rangle (3m_0^2-2M^2) WF[\mathcal{T}_2,\bar v]
 +216 m_0^2 864 M^4\langle \bar qq\rangle \Big(5e_d WF[\tilde S,v]+2e_u WF[\tilde S,\bar v]\Big)\nonumber\\
 & +3456 (e_d-e_u)f_{3\gamma}m_0^2m_c~M^4 WF[\psi^\nu,u]
 +(4e_u-e_d)\langle g_s^2 G^2 \rangle \langle \bar qq \rangle (2m_0^2-5M^2)WFD[\mathbb{A},u]\nonumber\\
 &-216~ e_d f_{3\gamma} m_0^2m_c~M^4\Big(WFD[\mathcal{A},v]-WFD[\mathcal{V},v]\Big)
 -216~ e_u f_{3\gamma} m_0^2m_c~M^4\Big(WFD[\mathcal{A},\bar v]-WFD[\mathcal{V},\bar v]\Big)\nonumber\\
 &-2(4e_u-e_d)\chi \langle g_s^2 G^2\rangle\langle \bar qq\rangle m_0^2 M^2
 WFD[\varphi_\gamma,u]\Bigg)\Bigg]FlNP[2,4,1]\nonumber\\
 &+\frac{m_c^3}{497664 M^{12} \pi^4}\Bigg[
 18(e_d-e_u)\pi^2 f_{3\gamma}\langle \bar qq \rangle M^4(4M^2-5m_0^2)\psi^a(u_0) 
 +\langle \bar qq \rangle \pi^2\Bigg(2592(e_u-e_d)f_{3\gamma}m_0^2 M^4 \psi^\nu(u_0)\nonumber\\
 &-324 e_d \langle \bar qq \rangle M^4(45 m_0^2-32 M^2)WF[\mathcal{T}_1,v]
 -324 e_u \langle \bar qq \rangle M^4(12m_0^2-8 M^2)WF[\mathcal{T}_1,\bar v]\nonumber\\
 &-324 e_d \langle \bar qq \rangle M^4(15 m_0^2-8 M^2)WF[\mathcal{T}_2,v]
 -324 e_u \langle \bar qq \rangle  M^4(12m_0^2-8 M^2)WF[\mathcal{T}_2,\bar v] \nonumber\\
 &-324 e_u \langle \bar qq \rangle  M^4\Big(5~e_d m_0^2 WF[\tilde S, v]
 -2~e_u~M^2 WF[\tilde S, \bar v]\Big)
 +5184(e_u-e_d) f_{3\gamma}m_c~m_0^2 M^4 WFD[\psi^\nu,u]\nonumber\\
 &+ (4 e_u-e_d)\langle g_s^2G^2 \rangle \langle \bar qq \rangle(5m_0^2-2M^2)WFD[\mathbb{A},u]
 +108 e_d f_{3\gamma}m_0^2 M^2 m_c\Big(2 WFD[\mathbb{A},v]+WFD[\mathcal{V},v]\Big)\nonumber\\
 &+108 e_u f_{3\gamma}m_0^2 M^2 m_c\Big(2 WFD[\mathbb{A},\bar v]+WFD[\mathcal{V},\bar v]\Big)
 -2(4e_u-e_d)\chi \langle g_s^2G^2 \rangle \langle \bar qq \rangle m_0^2 M^2
 WFD[\varphi_\gamma,u]\Bigg)\Bigg]\nonumber\\
 &FlNP[1,4,1]\nonumber\\
 &+\frac{m_c^5}{248832 M^{12} \pi^4}\Bigg[-36(e_u-e_d)f_{3\gamma}M^4\Big(7 \langle g_s^2G^2 \rangle M^2
 +144~m_c~\pi^2\langle \bar qq \rangle(4M^2-5m_0^2)\Big)\psi^a(u_0)\nonumber\\
 &+\pi^2\langle \bar qq \rangle\Bigg(5184(e_d-e_u)f_{3\gamma}m_0^2 m_c~M^2 \psi^\nu(u_0)
 +648~e_d\langle \bar qq \rangle M^4(45m_0^2-32 M^2)WF[\mathcal{T}_1,v]\nonumber\\
 &+2592~e_u\langle \bar qq \rangle M^4(3m_0^2-2 M^2)WF[\mathcal{T}_1,\bar v]
 +2592~e_d\langle \bar qq \rangle M^4(4m_0^2-2 M^2)WF[\mathcal{T}_2, v]\nonumber\\
 &+2592~e_u\langle \bar qq \rangle M^4(3m_0^2-2 M^2)WF[\mathcal{T}_2,\bar v]
 +648 \langle \bar qq \rangle M^4\Big(5~e_d~m_0^2 WF[\tilde S,v]+2~e_u~M^2 WF[\tilde S, \bar v]\Big)\nonumber\\
 & +(4 e_u-e_d)\langle g_s^2G^2 \rangle \langle \bar qq \rangle(2M^2-5m_0^2)WFD[\mathbb{A},u]
 -216 e_d f_{3\gamma}m_0^2 M^4 m_c\Big(WFD[\mathbb{A},v]+WFD[\mathcal{V},v]\Big)\nonumber\\
 &-216 e_u f_{3\gamma}m_0^2 M^4 m_c\Big(WFD[\mathbb{A},\bar v]+WFD[\mathcal{V},\bar v]\Big)
 +2(4e_u-e_d)\chi \langle g_s^2G^2 \rangle \langle \bar qq \rangle m_0^2 M^2
 WFD[\varphi_\gamma,u]\Bigg)\Bigg]FlNP[0,4,1]\nonumber
\end{align}

\begin{align}
&+\frac{\langle g_s^2 G^2 \rangle \langle \bar qq \rangle}{1327104 M^{10} \pi^4}\Bigg[
2(e_u+e_d)M^4 A(u_0)+3(e_u-e_d)\pi^2m_0^2f_{3\gamma}WFD[\psi^a,u]\Bigg]FlNP[3,3,1]\nonumber\\
 &+\frac{ m_c \langle \bar qq \rangle}{1990656 M^{10} \pi^4}\Bigg[
 (e_u-e_d)\Bigg(-3456\pi^2m_c^2 M^2\langle \bar qq \rangle (3m_0^2-2M^2)
+\langle g_s^2 G^2 \rangle \Big(-21 m_c M^4\nonumber\\
&+20\pi^2 \langle \bar qq \rangle(16m_0^2-11M^2)\Big)\Bigg)A(u_0)
+4\pi^2\Bigg\{(e_d-e_u)\chi M^2 \langle \bar qq \rangle \Big(5 \langle g_s^2 G^2 \rangle (11m_0^2-8M^2)
+1728 m_0^2 m_c^2 M^2\Big)\varphi_\gamma(u_0)\nonumber\\
&+4(e_u+e_d)\langle \bar qq \rangle \Big(7\langle g_s^2 G^2 \rangle (2m_0^2-M^2)
-216  m_c^2 m_0^2 M^2\Big)WF[\varphi_\gamma,u]
+9(e_d-e_u)m_c m_0^2 f_{3\gamma}\langle g_s^2 G^2 \rangle WF[\psi^a,u]\Bigg\}\Bigg]\nonumber\\
&FlNP[2,3,1]\nonumber\\
&+\frac{m_c^3 \langle \bar qq \rangle }{248832 M^{10} \pi^4}\Bigg[
\Bigg(3(25e_u-13e_d)m_c~M^4\langle g_s^2 G^2 \rangle -4 (e_u-e_d)\pi^2 \langle \bar qq \rangle\Big(
-864~\pi^2 m_c^2 \langle \bar qq \rangle (3 m_0^2-2M^2)\nonumber\\
&-\langle g_s^2 G^2 \rangle(80 m_0^2+55 M^2)\Big)\Bigg)A(u_0)
+4(e_d-e_u) \pi^2 \chi \langle \bar qq \rangle M^2 \Big(5\langle g_s^2 G^2 \rangle(11m_0^2-8M^2)+
1728m_0^2 m_c^2 M^2\Big)\varphi_\gamma\nonumber\\
&+4 \pi^2 \langle \bar qq \rangle\Big((2e_d-7e_u)\langle g_s^2 G^2 \rangle(M^2-2m_0^2)
+216(e_d-e_u)m_0^2m_c^2 M^2\Big)WF[h_\gamma,u]\nonumber\\
&+9(e_d-e_u)\pi^2 f_{3\gamma}\langle g_s^2 G^2 \rangle m_0^2 m_c~WF[\psi^a,u]\Bigg]FlNP[1,3,1]\nonumber\\
&+\frac{m_c^5}{82944 M^{10}\pi^4}\Bigg[\langle \bar qq \rangle \Bigg\{
e_u\Bigg(2304\pi^2 m_c^2 M^2 \langle \bar qq \rangle(-3m_0^2+4M^2)
+\langle g_s^2 G^2 \rangle \Big(-73m_c M^4+120\pi^2 \langle \bar qq \rangle (m_0^2-M^2)\Big)\Bigg)\nonumber\\
&+e_d\Bigg(2304\pi^2 m_c^2 M^2 \langle \bar qq \rangle(3m_0^2-4M^2)
+\langle g_s^2 G^2 \rangle \Big(99m_c M^4+120\pi^2 \langle \bar qq \rangle (m_0^2M^2)\Big)\Bigg)
\Bigg\}A(u_0)\nonumber\\
&-8M^2 \chi \langle \bar qq \rangle\Bigg\{
e_u\Bigg(1152\pi^2 m_c^2 M^2 \langle \bar qq \rangle(m_0^2-2M^2)
+\langle g_s^2G^2 \rangle\Big(23m_c M^4+5\pi^2 \langle \bar qq \rangle(-3m_0^2+4M^2)\Big)\Bigg)\nonumber\\
&+e_d\Bigg(-1152\pi^2 m_c^2 M^2 \langle \bar qq \rangle(m_0^2-M^2)
+5\langle g_s^2G^2 \rangle\Big(-3m_c M^4+\pi^2 \langle \bar qq \rangle(3m_0^2-4M^2)\Big)\Bigg)
\Bigg\}\varphi_\gamma(u_0)\nonumber\\
&+\langle \bar qq \rangle\Bigg\{e_u\Bigg(576\pi^2 m_c^2 M^2 \langle \bar qq \rangle(3m_0^2-4M^2)
+\langle g_s^2 G^2 \rangle\Big(17m_cM^4+56\pi^2\langle \bar qq \rangle(-m_0^2+M^2)\Big)\Bigg)\nonumber\\
&+e_d\Bigg(288\pi^2 m_c^2 M^2 \langle \bar qq \rangle(-3m_0^2+4M^2)
+\langle g_s^2 G^2 \rangle\Big(-13m_cM^4+8\pi^2\langle \bar qq \rangle(-m_0^2+M^2)\Big)\Bigg)
\Bigg\}WF[h_\gamma,u]\nonumber\\
&+4(e_u-e_d)M^6 f_{3\gamma}\langle g_s^2 G^2 \rangle \psi^a(u_0)\Bigg]FlNP[2,3,0]\nonumber\\
\end{align}

\begin{align}
&+\frac{m_c^3}{165888 M^{10}\pi^4}\Bigg[3\langle \bar qq \rangle \Bigg\{
e_u\Bigg(768\pi^2 m_c^2 M^2 \langle \bar qq \rangle(-3m_0^2+4M^2)
+\langle g_s^2 G^2 \rangle \Big(23m_c M^4+40\pi^2 \langle \bar qq \rangle (m_0^2-M^2)\Big)\Bigg)\nonumber\\
&+e_d\Bigg(768\pi^2 m_c^2 M^2 \langle \bar qq \rangle(3m_0^2-4M^2)
+\langle g_s^2 G^2 \rangle \Big(31m_c M^4+40\pi^2 \langle \bar qq \rangle (m_0^2-M^2)\Big)\Bigg)
\Bigg\}A(u_0)\nonumber\\
&-8M^2 \chi \langle \bar qq \rangle\Bigg\{
e_u\Bigg(1152\pi^2 m_c^2 M^2 \langle \bar qq \rangle(m_0^2-2M^2)
+\langle g_s^2G^2 \rangle\Big(25m_c M^4+5\pi^2 \langle \bar qq \rangle(-3m_0^2+4M^2)\Big)\Bigg)\nonumber\\
&+e_d\Bigg(-1152\pi^2 m_c^2 M^2 \langle \bar qq \rangle(m_0^2-M^2)
+\langle g_s^2G^2 \rangle\Big(-13m_c M^4+5\pi^2 \langle \bar qq \rangle(3m_0^2-4M^2)\Big)\Bigg)
\Bigg\}\varphi_\gamma(u_0)\nonumber\\
&+4\langle \bar qq \rangle\Bigg\{e_u\Bigg(576\pi^2 m_c^2 M^2 \langle \bar qq \rangle(3m_0^2-4M^2)
+\langle g_s^2 G^2 \rangle\Big(17m_cM^4+56\pi^2\langle \bar qq \rangle(-m_0^2+M^2)\Big)\Bigg)\nonumber\\
&+e_d\Bigg(288\pi^2 m_c^2 M^2 \langle \bar qq \rangle(-3m_0^2+4M^2)
+\langle g_s^2 G^2 \rangle\Big(-13m_cM^4+8\pi^2\langle \bar qq \rangle(-m_0^2+M^2)\Big)\Bigg)
\Bigg\}WF[h_\gamma,u]\nonumber\\
&-32(e_u-e_d)M^6 f_{3\gamma}\langle g_s^2 G^2 \rangle \psi^a(u_0)\Bigg]FlNP[1,3,0]\nonumber\\
 &+\frac{m_c}{1327104 M^{10}\pi^4}\Bigg[3\langle \bar qq \rangle \Bigg\{
e_u\Bigg(768\pi^2 m_c^2 M^2 \langle \bar qq \rangle(-3m_0^2+4M^2)
+\langle g_s^2 G^2 \rangle \Big(35m_c M^4+40\pi^2 \langle \bar qq \rangle (m_0^2-M^2)\Big)\Bigg)\nonumber\\
&+e_d\Bigg(-768\pi^2 m_c^2 M^2 \langle \bar qq \rangle(3m_0^2-4M^2)
+\langle g_s^2 G^2 \rangle \Big(-19m_c M^4+40\pi^2 \langle \bar qq \rangle (m_0^2-M^2)\Big)\Bigg)
\Bigg\}A(u_0)\nonumber\\
&-8M^2 \chi \langle \bar qq \rangle\Bigg\{
e_u\Bigg(1152\pi^2 m_c^2 M^2 \langle \bar qq \rangle(m_0^2-2M^2)
+\langle g_s^2G^2 \rangle\Big(31m_c M^4+5\pi^2 \langle \bar qq \rangle(-3m_0^2+4M^2)\Big)\Bigg)\nonumber\\
&+e_d\Bigg(-1152\pi^2 m_c^2 M^2 \langle \bar qq \rangle(m_0^2-M^2)
+\langle g_s^2G^2 \rangle\Big(-7m_c M^4+5\pi^2 \langle \bar qq \rangle(3m_0^2-4M^2)\Big)\Bigg)
\Bigg\}\varphi_\gamma(u_0)\nonumber\\
&+4\langle \bar qq \rangle\Bigg\{e_u\Bigg(576\pi^2 m_c^2 M^2 \langle \bar qq \rangle(3m_0^2-4M^2)
+\langle g_s^2 G^2 \rangle\Big(17m_cM^4+56\pi^2\langle \bar qq \rangle(-m_0^2+M^2)\Big)\Bigg)\nonumber\\
&+e_d\Bigg(288\pi^2 m_c^2 M^2 \langle \bar qq \rangle(-3m_0^2+4M^2)
+\langle g_s^2 G^2 \rangle\Big(-13m_cM^4+8\pi^2\langle \bar qq \rangle(-m_0^2+M^2)\Big)\Bigg)
\Bigg\}WF[h_\gamma,u]\nonumber\\
&-24(e_u-e_d)M^6 f_{3\gamma}\langle g_s^2 G^2 \rangle \psi^a(u_0)\Bigg]FlNP[0,3,0]\nonumber\\
 &+\frac{\langle g_s^2 G^2 \rangle \langle \bar qq \rangle}{663552 M^{10} \pi^4}\Bigg[
 (e_u+e_d)M^4\Big(3M^2\chi \varphi_\gamma(u_o)-A(u_0)\Big)
 -6(e_u-e_d)\pi^2 f_{3\gamma}(m_0^2-M^2)WFD[\psi^a,u]\Bigg]FlNP[-1,3,0]
 \nonumber\\
 &+\frac{m_c^3}{3317760 M^{10} \pi^4}\Bigg[
 -3456(e_d-e_u)f_{3\gamma} M^6 \Big(FlNP[1,5,0]
 -12~m_c^2 FlNP[2,5,0]+48~m_c^4 FlNP[3,5,0]\Big)\psi^a(u_0)\nonumber\\ 
 &+144 \pi^2 \langle \bar qq \rangle^2(m_0^2-M^2)\Bigg\{
 e_u\Big(-2 WF[\mathcal T_1,\bar v]-2 WF[\mathcal T_2,\bar v]+2WF[\tilde S,\bar v]
 + WFD[\mathcal T_1,\bar v]+ WFD[\mathcal T_2,\bar v]+ WFD[\tilde S,\bar v]\Big)\nonumber\\
 &+e_d\Big(-8 WF[\mathcal T_1,v]
  -2 WF[\mathcal T_2,v]+2e_d WF[\tilde S,v]+4 WFD[\mathcal T_1,v]
 + WFD[\mathcal T_2,v]-e_d WFD[\tilde S,v]\Big)\Bigg\} \nonumber\\
 & +(e_u-e_d)f_{3\gamma}\Big( 7 \langle g_s^2 G^2 \rangle M^2
 +576 \langle \bar qq \rangle m_c(M^2-m_0^2)\Big)WFD[\psi^a,u]\Bigg]\Bigg(FlNP[0,5,0]-12~m_c^2FlNP[1,5,0]\nonumber\\
 &+48~m_c^4FlNP[2,5,0]-64~m_c^6 FlNP[3,5,0]\Bigg).
\end{align}

 \begin{align} 
  &\Pi_3 =0.\\
 \nonumber\\
 \nonumber\\
 &\Pi_4=-\frac{m_c^3 M^2 \chi \langle g_s^2G^2\rangle \langle \bar qq \rangle  }{18432 \pi^4}
(e_u+e_d) \varphi_\gamma(u_0)N[1,3,2]\nonumber\\
&+\frac{5 m_c^2 f_{3\gamma}\langle g_s^2G^2\rangle }{6912\pi^4}
(e_u-e_d)WF[\psi^\nu,u]~N[1,1,0]\nonumber\\
&+\frac{f_{3\gamma}m_c^4}{16\pi^4}WF[\psi^\nu,u]~N[3,3,1]\nonumber\\
&+ \frac{f_{3\gamma}\langle g_s^2G^2\rangle m_c^2 M^2 }{884736 \pi^4}
WF[\psi^\nu,u]~N[1,1,1]\nonumber\\
&+\frac{m_c^2}{2304 \pi^4}\Bigg[-(e_u+e_d) \chi \langle g_s^2G^2\rangle \langle \bar qq \rangle \varphi_\gamma(u_0)
+18m_cM^2f_{3\gamma} \Big(2e_u WF[\mathcal A,\bar v]+5e_d WF[\mathcal A,v]\Big)\Bigg]N[1,2,0]\nonumber\\
&+\frac{m_c^3 \langle \bar qq \rangle }{128 \pi^4}\Bigg[
e_d\Big(5WF[\mathcal T_1,v]+5WF[\mathcal T_2,v]+3WF[\tilde S,v]\Big)
+2e_u\Big(WF[\mathcal T_1,\bar v]+WF[\mathcal T_2,\bar v]\Big)\Bigg]\nonumber\\
&\Bigg(3~m_c~N[2,3,1]-4~N[1,3,0]+4~m_c~N[1,4,1]\Bigg)\nonumber\\
&+\frac{m_c^3 }{2304 \pi^4}\Bigg[(e_u+e_d)\chi \langle g_s^2G^2\rangle \langle \bar qq \rangle \varphi_\gamma(u_0)
-9m_cM^2 f_{3\gamma}\Big((5~e_d WF[\mathcal A,v]+2~e_u~WF[\mathcal A,\bar v])\Big)\nonumber\\
&+9~M^2 \langle \bar qq \rangle\Bigg(
e_d\Big(5WF[\mathcal T_1,v]+5WF[\mathcal T_2,v]+3WF[\tilde S,v]\Big)
+2e_u\Big(WF[\mathcal T_1,\bar v]+WF[\mathcal T_2,\bar v]\Big)\Bigg)\Bigg]N[1,3,1]\nonumber\\
&-\frac{m_c^4 M^2 \langle \bar qq \rangle }{1024 \pi^4}\Bigg[
e_d\Big(5WF[\mathcal T_1,v]+5WF[\mathcal T_2,v]+3WF[\tilde S,v]\Big)
+2e_u\Big(WF[\mathcal T_1,\bar v]+WF[\mathcal T_2,\bar v]\Big)\Bigg]\nonumber\\
&\Bigg(4~N[1,4,2]+3~N[2,3,2]\Bigg)\nonumber\\
&-\frac{m_c^2 \langle g_s^2G^2\rangle }{221184 \pi^2}\Bigg[
\langle \bar qq \rangle \Bigg(6(e_u+e_d)\Big(A(u_0)+2\chi M^2\varphi_\gamma(u_0)\Big)
+22e_d\Big(2WF[\mathcal T_1,v]+5WF[\mathcal T_2,v]-WF[\mathcal T_3,v]\nonumber\\
&+2WF[\mathcal T_4,v]\Big)+ e_u\Big(3WF[\mathcal S,\bar v]+44WF[\mathcal T_1,\bar v]
+113WF[\mathcal T_2,\bar v]-25WF[\mathcal T_3,\bar v]+44WF[\mathcal T_4,\bar v]\Big)\Bigg)\nonumber\\
&-48(e_u+e_d)\langle \bar qq \rangle WF[f_\gamma,u]
-120(e_u-e_d)f_{3\gamma} m_c~WF[\psi^\nu,u]\Bigg]N[1,2,1]\nonumber\\
&-\frac{m_c^2 M^2 \langle g_s^2G^2\rangle }{1769472 \pi^2}\Bigg[
6(e_u+e_d)\langle \bar qq \rangle A(u_0)
+22e_d \langle \bar qq \rangle\Big(2WF[\mathcal T_1,v]+5WF[\mathcal T_2,v]-WF[\mathcal T_3,v]\nonumber\\
&+2WF[\mathcal T_4,v]\Big)+ e_u \langle \bar qq \rangle \Big(3WF[\mathcal S,\bar v]-44WF[\mathcal T_1,\bar v]
+113WF[\mathcal T_2,\bar v]-25WF[\mathcal T_3,\bar v]+44WF[\mathcal T_4,\bar v]\Big)\Bigg)\nonumber\\
&-48(e_u+e_d)\langle \bar qq \rangle WF[f_\gamma,u]
-280(e_u-e_d)f_{3\gamma} m_c~WF[\psi^\nu,u]\Bigg]N[1,2,2]\nonumber\\
&+\frac{11 f_{3\gamma}\langle g_s^2G^2\rangle m_c^2}{9216 M^2 \pi^4}\Bigg[
 e_u \Big( 2WF[\mathcal A,\bar v]-WFD[\mathcal A,\bar v]\Big)
+e_d\Big(2 WF[\mathcal A,v]- WFD[\mathcal A,v]\Big) \Bigg]N[2,2,0]\nonumber\\
 &+\frac{m_c^2}{110592 \pi^4}\Bigg[36(e_u+e_d)
 m_c \chi \langle g_s^2G^2\rangle \langle \bar qq \rangle  \varphi_\gamma(u_0)
 +f_{3\gamma}\Bigg(4e_d\Big(11\langle g_s^2G^2\rangle+540 m_c^2 M^2\Big) WF[\mathcal A,v]\nonumber\\
 &+e_u\Big(44 \langle g_s^2G^2\rangle+864 m_c^2M^2\Big)WF[\mathcal A,\bar v]
 -33 \langle g_s^2G^2\rangle\Big(e_d WFD[\mathcal A,v]
 +e_u WFD[\mathcal A,\bar v]\Big)\Bigg)\Bigg]N[2,2,1]\nonumber
\end{align}

\begin{align}
 &+\frac{m_c^2M^2\langle g_s^2G^2\rangle }{442368 \pi^4}\Bigg[
 -18(e_u+e_d) m_c \chi  \langle \bar qq \rangle  \varphi_\gamma(u_0)
 +11 f_{3\gamma}\Bigg(e_d\Big( WF[\mathcal A,v]-WFD[\mathcal A,v]\Big)+e_u \Big( 2WF[\mathcal A,\bar v]\nonumber\\
&- WFD[\mathcal A,\bar v]\Big)\Bigg)\Bigg]N[2,2,2]\nonumber\\
&+\frac{ 11m_c^2 M^2 f_{3\gamma} \langle g_s^2G^2\rangle }{3538944 \pi^4}\Bigg[
  +e_d\Big( 2WF[\mathcal A,v]-WFD[\mathcal A,v]\Big)+e_u \Big( WF[\mathcal A,\bar v]+
- WFD[\mathcal A,\bar v]\Big)\Bigg]N[2,2,3]\nonumber\\
&-\frac{m_cm_0^2\langle \bar qq \rangle ^2 }{4608 M^{10}\pi^2}\Bigg[
  5e_d\Big(WF[\mathcal T_1,v]+WF[\mathcal T_2,v]\Big)
  +2e_u\Big(WF[\mathcal T_1,\bar v]+WF[\mathcal T_2,\bar v]\Big)\Bigg]
  \Bigg(64~m_c^6 FlNP[1,4,2]\nonumber\\
  &-48~m_c^4FlNP[2,4,2]+12m_c^2 FlNP[3,4,3]-FlNP[4,4,2]\Bigg)\nonumber\\
  &-\frac{m_c m_0^2\langle g_s^2G^2\rangle \langle \bar qq \rangle^2 }{331776 M^{12} \pi^2}
  (e_u+e_d)\Big(A(u_0)-8 WF[h_\gamma,u]\Big)
  \Bigg(16~m_c^4 FlNP[2,3,2]-8~m_c^2FlNP[3,3,2]\nonumber\\
  &+FlNP[4,3,2]\Bigg)\nonumber\\
  &+\frac{m_c \langle \bar qq \rangle }{165888 M^{12}\pi^2}\Bigg[
  (e_u+e_d)\langle g_s^2G^2 \rangle \langle \bar qq \rangle(5m_0^2-2M^2)A(u_0)
  +2m_0^2 M^2\Bigg((e_u+e_d)\chi \langle g_s^2G^2 \rangle \langle \bar qq \rangle \varphi_\gamma(u_0)\nonumber\\
 & +36m_cM^2 f_{3\gamma}\Big(e_d WF[\mathcal A,v]-2e_u FW[\mathcal A, \bar v]\Big)\Bigg)
 -8(e_u+e_d)\langle g_s^2G^2 \rangle \langle \bar qq \rangle (5m_0^2-2M^2)WF[h_\gamma,u]\Bigg]\nonumber\\
 &\Bigg(16~m_c^4 FlNP[0,3,1]-8~m_c^2 FlNP[1,3,1]+FlNP[2,3,1]\Bigg)\nonumber\\
& +\frac{m_c}{663552 M^{12} \pi^4}\Bigg[
 (e_u+e_d)\langle g_s^2G^2 \rangle \langle \bar qq \rangle\Big(-3m_c M^4
 -8\pi^2\langle \bar qq \rangle (5m_0^2-4M^2)\Big)A(u_0) \nonumber\\
 &+8~M^2\Bigg\{4(e_u+e_d)\pi^2 \chi \langle g_s^2G^2 \rangle \langle \bar qq \rangle^2(m_0^2-M^2)\varphi_\gamma(u_0)
 +M^2f_{3\gamma}\Bigg(e_d\Big(17 \langle g_s^2G^2 \rangle M^2
 +36 \pi^2 m_c \langle \bar qq \rangle(3m_0^2-4M^2)\Big)\nonumber\\
 &WF[\mathcal A,v]+e_u\Big(17 \langle g_s^2G^2 \rangle M^2
 -72 \pi^2m_c\langle \bar qq \rangle(3m_0^2-4M^2)\Big)WF[\mathcal A,\bar v]\Bigg)\Bigg\}
 +8(e_u+e_d)\langle g_s^2G^2 \rangle \langle \bar qq \rangle\Big(3~m_c~M^4\nonumber\\
 &+8\pi^2 \langle \bar qq \rangle(5m_0^2-4M^2)\Big)WF[h_\gamma,u]\Bigg]
 \Bigg(16~m_c^4FlNP[2,3,0]-8~m_c^2 FlNP[1,3,0]+FlNP[0,3,0]\Bigg)\nonumber\\
 &-\frac{m_0^2 \langle \bar qq \rangle }{165888 M^{10} \pi^2}\Bigg[
 432 m_c M^2 \langle \bar qq \rangle \Bigg(
 e_d\Big(WF[\mathcal T_1,v]+WF[\mathcal T_2,v]-WF[\mathcal T_3,v]-WF[\mathcal T_4,v]\Big)
 +e_u\Big(WF[\mathcal T_1,\bar v]\nonumber\\
 &+WF[\mathcal T_2,\bar v]-WF[\mathcal T_3,\bar v]-WF[\mathcal T_4,\bar v]\Big)
 -5(e_u-e_d)f_{3\gamma}\langle g_s^2G^2 \rangle WF[\psi^\nu,u]\Bigg]
 \Bigg(16~m_c^24 FlNP[1,2,1]\nonumber\\
 &+8~m_c^2 FlNP[2,2,1]-FlNP[3,2,1]\Bigg)\nonumber
\end{align}

\begin{align}
& +\frac{m_c \langle \bar qq \rangle}{9216 M^{10}\pi^2}\Bigg[\langle \bar qq \rangle(17m_0^2-8M^2)\Bigg(
5e_d\Big(WF[\mathcal T_1,v]+WF[\mathcal T_2,v]\Big)
+2e_u\Big(WF[\mathcal T_1,\bar v]+WF[\mathcal T_2,\bar v]\Big)\bigg)\nonumber\\
&+3e_d \langle \bar qq \rangle m_0^2 WF[\tilde S,v]
-32(e_u-e_d)f_{3\gamma}m_c~m_0^2 WF[\psi^\nu,u]\bigg]
\Bigg(64~m_c^6FlNP[-1,4,1]-48~m_c^4FlNP[0,4,1]\nonumber\\
&-12~m_c^2 FLNP[1,4,1]-FlNP[2,4,1]\Bigg)\nonumber\\
 &+\frac{m_c}{27648 M^{10}\pi^4}\Bigg[ \langle \bar qq \rangle\Bigg(
 3~m_c~M^4+28\pi^2\langle \bar qq \rangle (m_0^2-M^2)\Bigg)
 \Bigg(15e_d\Big(WF[\mathcal T_1,v]+WF[\mathcal T_2,v]\Big)\nonumber\\
&+6e_u\Big(WF[\mathcal T_1,\bar v]+WF[\mathcal T_2,\bar v]\Big)\Bigg)
+12 e_d \pi^2 \langle \bar qq \rangle^2(m_0^2-M^2)WF[\tilde S,v]
+4(e_u-e_d)f_{3\gamma}\Big(\langle g_s^2G^2 \rangle M^2\nonumber\\
&+96 \pi^2 \langle \bar qq \rangle m_c(-m_0^2+M^2)\Big) WF[\psi^\nu, u]\Bigg]
\Bigg(-64~m_c^6 FlNP[3,4,0]+48~m_c^4 FlNP[2,4,0]-12~m_c^2 FlNP[1,4,0]\nonumber\\
&+FlNP[0,4,0]\Bigg)\nonumber\\
&-\frac{\langle \bar qq \rangle}{165888 M^{10}\pi^4}\Bigg[
3e_d M^2\Big(23\langle g_s^2G^2 \rangle M^2 + 288\pi^2 \langle \bar qq \rangle m_c(4M^2-3m_0^2)\Big)WF[\mathcal T_1,v]
+6e_u \Big(17\langle g_s^2G^2 \rangle M^2 \nonumber\\
&+ 144\pi^2 \langle \bar qq \rangle m_c(4M^2-3m_0^2)\Big)WF[\mathcal T_1,\bar v]
+e_d M^2\Big(69\langle g_s^2G^2 \rangle M^2 
+ 864\pi^2 \langle \bar qq \rangle m_c(4M^2-3m_0^2)\Big)WF[\mathcal T_2,v]\nonumber\\
&+e_u \Big(102 \langle g_s^2G^2 \rangle M^2 
+ 864\pi^2 \langle \bar qq \rangle m_c(4M^2-3m_0^2)\Big)WF[\mathcal T_2,\bar v]
-e_d M^2\Big(36\langle g_s^2G^2 \rangle M^2 \nonumber\\
&+ 864\pi^2 \langle \bar qq \rangle m_c(4M^2-3m_0^2)\Big)WF[\mathcal T_3,v]
-e_u \Big(36 \langle g_s^2G^2 \rangle M^2 
+ 864\pi^2 \langle \bar qq \rangle m_c(4M^2-3m_0^2)\Big)WF[\mathcal T_3,\bar v]\nonumber\\
&-e_d M^2\Big(36\langle g_s^2G^2 \rangle M^2 
+ 864\pi^2 \langle \bar qq \rangle m_c(4M^2-3m_0^2)\Big)WF[\mathcal T_4,v]
-e_u \Big(36 \langle g_s^2G^2 \rangle M^2 \nonumber\\
&+ 864\pi^2 \langle \bar qq \rangle m_c(4M^2-3m_0^2)\Big)WF[\mathcal T_4,\bar v]
-40(e_u-e_d) \pi^2 f_{3\gamma} \langle g_s^2G^2 \rangle(m_0^2-M^2)WF[\psi^\nu,u]\Bigg]
\Bigg(16~m_c^4 FlNP[1,2,0]\nonumber\\
&-8~m_c^2 FlNP[0,2,0]+FlNP[-1,2,0]\Bigg).
\end{align}

The functions $N[n,m,k]$,~$FlP[n,m,k]$,~$FlNP[n,m,k]$,~$WFD[\mathcal{A},\bar v]$,~$WFD[\mathcal{A}, v]$,~$WF[\mathcal{A},\bar v]$, ~$WF[\mathcal{A}, v]$, ~$WFD[\mathcal{A},u]$ and $WF[\mathcal{A},u]$ are
defined as:
\begin{align}
 N[n,m,k]&=\int_0^\infty dt\int_0^\infty dt'~ 
 \frac{e^{-m_c/2(t+t')}}{t^n~(\frac{m_c}{t}+\frac{m_c}{t'})^k~ t'^m },\nonumber\\
 FlP[n,m,k]&= \int_{4m_c^2}^{s_0} ds \int_{4m_c^2}^s dl~ \frac{e^{-l^2/\phi}~ l^n~ (l-s)^m}{(4m^2-l)^2~\phi^k},\nonumber\\
 FlNP[n,m,k]&=  \int_{4m_c^2}^{s_0} ds \int_{4m_c^2}^s dl~ \frac{e^{-l^2/\beta}~ l^n~ (l-s)^m}{(l-2m_c^2)~\beta^k},\nonumber\\
 WFD[\mathcal{A},\bar v]&=\int D_{\alpha_i} \int_0^1 dv~ \mathcal{A}(\alpha_{\bar q},\alpha_q,\alpha_g)
 \delta'(\alpha_ q +\bar v \alpha_g-u_0),\nonumber\\
  WFD[\mathcal{A}, v]&=\int D_{\alpha_i} \int_0^1 dv~ \mathcal{A}(\alpha_{\bar q},\alpha_q,\alpha_g)
 \delta'(\alpha_{\bar q}+ v \alpha_g-u_0),\nonumber\\
  WF[\mathcal{A},\bar v]&=\int D_{\alpha_i} \int_0^1 dv~ \mathcal{A}(\alpha_{\bar q},\alpha_q,\alpha_g)
 \delta(\alpha_ q +\bar v \alpha_g-u_0),\nonumber
 \end{align}
 \begin{align}
  WF[\mathcal{A}, v]&=\int D_{\alpha_i} \int_0^1 dv~ \mathcal{A}(\alpha_{\bar q},\alpha_q,\alpha_g)
 \delta(\alpha_{\bar q}+ v \alpha_g-u_0),\nonumber\\
 WFD[\mathcal{A},u]&=\int_0^1 du~ A(u)\delta'(u-u_0),\nonumber\\
 WF[\mathcal{A},u]&=\int_0^1 du~ A(u),\nonumber\\
 \end{align}
 where
\begin{align}
 \beta=4~l~M^2-16m_c^2M^2,~~~~~~~~~~~~~~~~~\phi=8~l~M^2-32m_c^2M^2.\nonumber
\end{align}
\section*{Appendix C:}
In this appendix, we give some details on Fourier and  Borel transformations as well as continuum subtraction.
We take a term in the form

 \begin{align}
  I=\int_0^1 du A(u)\int d^4x e^{i(p+q u )x} \frac{K_\nu(m_Q\sqrt{-x^2})}{\sqrt{-x^2}^\nu}
  \frac{K_\mu(m_Q\sqrt{-x^2})}{\sqrt{-x^2}^\mu},
 \end{align}
where $K_{\nu}$ comes
from the heavy quark propagator. To proceed we apply
the integral representation of the Bessel function of second kind as
 \begin{equation*}
\frac{K_{\nu }\left( m_{Q}\sqrt{-x^{2}}\right) }{\left( \sqrt{-x^{2}}\right)
^{\upsilon }}=\frac{1}{2}\int_{0}^{\infty }\frac{dt}{t^{\nu +1}}\exp \left[ -%
\frac{m_{Q}}{2}\left( t-\frac{x^{2}}{t}\right) \right].
\end{equation*}
As a result, we get
\begin{align}
  I=\int_0^1 du A(u)\int d^4x e^{i(p+q u )x} \int_{0}^{\infty }\frac{dt}{t^{\nu +1}}\exp \left[ -%
\frac{m_{Q}}{2}\left( t-\frac{x^{2}}{t}\right) \right] \int_{0}^{\infty }\frac{dt'}{t'^{\mu +1}}\exp \left[ -%
\frac{m_{Q}}{2}\left( t'-\frac{x^{2}}{t'}\right) \right].
 \end{align}
By applying the Wick rotation we obtain
\begin{align}
  I=\int_0^1 du A(u) \int_{0}^{\infty }\frac{dt}{t^{\nu +1}}
  \int_{0}^{\infty }\frac{dt'}{t'^{\mu +1}}\exp\Bigg[{-\frac{m_Q }{2}(t+t')}\Bigg]
  \int d^4x \exp\Bigg[-i(p.x+q.x)-ax^2\Bigg],
 \end{align}
 
where $a= (\frac{m_Q}{t}+\frac{m_Q}{t'})$.
Taking the four-dimensional Gaussian integral we get 
\begin{align}
 I=\int_0^1 du A(u)\int_{0}^{\infty }\frac{dt}{t^{\nu +1}}
  \int_{0}^{\infty }\frac{dt'}{t'^{\mu +1}}\exp\Bigg[{-\frac{m_Q }{2}(t+t')}
 -\frac{(p+qu)^2}{4a}\Bigg]\frac{1}{a^2}.
\end{align}

Now, we apply the Borel transformation over the variables $p^2$ and $(p + q)^2$, which results in
\begin{align}
 I=\int_0^1 du A(u)\int_{0}^{\infty }\frac{dt}{t^{\nu +1}}
  \int_{0}^{\infty }\frac{dt'}{t'^{\mu +1}} \exp\Bigg[{-\frac{m_Q }{2}(t+t')}\Bigg]
  \frac{ M^2}{a^2} \delta\Big[\frac{1}{M^2}-\frac{1}{4a}\Big]~\delta\Big[u-u_0\Big].
\end{align}

After this step, we take the t integral using the corresponding Dirac delta. 
To do this, we use the property:
\begin{align}
 \delta(g(x)) =\frac{\delta(x-x_0)}{|g'(x)|}\theta(x_0),
\end{align}
and replace t by

\begin{align}
 t \rightarrow \Bigg(\frac{2m_Q~t'}{M^2 t'-2m_Q}\Bigg/ \Bigg|\frac{2~m_Q~t^2}{M^2t'-2~m_Q}\Bigg|\Bigg)
 ~\theta\Bigg(\frac{2m_Q~t'}{M^2 t'-2m_Q}\Bigg).
\end{align}

Then, we change the variable $t'\rightarrow s$ via

\begin{align}
 t' \rightarrow \frac{2~m_Q}{4~m_Q^2 M^2}~s.
\end{align}

Meanwhile, for the Borel transformations the following rules are applied:
 \begin{align} 
 &B_{p^2}B_{(p+q)^2}\exp\Bigg[-\frac{(p+q u)^2}{4a}\Bigg]~(p+q~u)^n~(p.q)^m \rightarrow M^2 ~(M^2/2)^m ~D\Big[\frac{1}{M^2},n\Big]
 \delta\Big[\frac{1}{M^2}-\frac{1}{4a}\Big]~\delta'\Big[u-u_0\Big],\nonumber\\
 & B_{p^2}B_{(p+q)^2}~\exp\Bigg[-\frac{(p+q u)^2}{4a}\Bigg]~ (p.q)^m \rightarrow M^2~ (M^2/2)^m~ 
 \delta\Big[\frac{1}{M^2}-\frac{1}{4a}\Big]~\delta'\Big[u-u_0\Big],\nonumber\\
 & B_{p^2}B_{(p+q)^2}~\exp\Bigg[-\frac{(p+q u)^2}{4a}\Bigg]~(p+q~u)^n \rightarrow M^2~ D\Big[\frac{1}{M^2},n\Big]
 \delta\Big[\frac{1}{M^2}-\frac{1}{4a}\Big],\nonumber\\
 & B_{p^2}B_{(p+q)^2}~\exp\Bigg[-\frac{(p+q u)^2}{4a}\Bigg]\rightarrow M^2 ~
 \delta\Big[\frac{1}{M^2}-\frac{1}{4a}\Big]~\delta\Big[u-u_0\Big].\nonumber\\
 \end{align}
 where, D represents the derivation and 
\begin{align}
M^2=\frac{M_1^2 M_2^2}{M_1^2+M_2^2},~~~~~~~~~
u_0=\frac{M_1^2}{M_1^2+M_2^2}.
\end{align}%

The following formula for the continuum subtraction is used
\begin{align}
&\left( M^{2}\right) ^{N}\int_{4m_Q^{2}}^{\infty }dse^{-s/M^{2}}f(s)\rightarrow
\int_{4m_Q^{2}}^{s_{0}}dse^{-s/M^{2}}F_{N}(s),
\end{align}%
where
\begin{align}
F_{N}(s)&= \Big(\frac{d}{ds}\Big)^{-N}f(s),~~~~~ N\leq 0,
\nonumber\\
F_{N}(s)&=\frac{1}{\Gamma (N)}\int_{4m_Q^{2}}^{s}dl~(s-l)^{N-1}f(l),~~~~~N>0,
\end{align}%
as a result of which we obtain the following expression:

\begin{align}
\int_0^1 du A(u) \int_{4 m_{Q'}}^{s_0} ds  \int_{4 m_{Q'}}^s dl~ 
\exp\Bigg[{-\frac{l+m_{Q'}\Big(-3-\frac{m_Q}{m_Q^2-m_{Q'}^2}\Big)}{M^2}}\Bigg]
\frac{(l-s)^3~\delta\Big[u-u_0\Big]}{3~m_Q~m^4_{Q'}~M^{12} 
\Bigg|\frac{m_Q\Big(-2m_Q+\frac{2m_{Q'}^2}{m_Q}\Big)^2}{m^4_{Q'}}\Bigg|},
\end{align}
with $ m_Q$ and $m_ {Q'} $ being the charm quark mass.
Here we face with  the well-known problem in the case of doubly heavy hadrons 
when  we take $ m_Q = m_ {Q'} $. The expression above becomes indeterminate. 
To get rid of this problem we take the limit of the expression in the integral, i.e.,
\begin{align}
\int_0^1 du A(u) \int_{4 m_{Q'}}^{s_0} ds  \int_{4 m_{Q'}}^s dl~
\lim_{{m_{Q'}}\rightarrow m_Q}\Bigg[
\exp\Bigg\{{-\frac{l+m_{Q'}\Big(-3-\frac{m_Q}{m_Q^2-m_{Q'}^2}\Big)}{M^2}}\Bigg\}
\frac{(l-s)^3 ~\delta\Big[u-u_0\Big]}{3~m_Q~m^4_{Q'}~M^{12} 
\Bigg|\frac{m_Q\Big(-2m_Q+\frac{2m_{Q'}^2}{m_Q}\Big)^2}{m^4_{Q'}}\Bigg|}\Bigg],
\end{align}
which gives a finite result.

\end{document}